\documentclass{article}
\usepackage[utf8]{inputenc}
\usepackage{comment}

\pdfoutput=1
\usepackage{amssymb}
\usepackage{amsmath}
\usepackage{setspace}
\usepackage{slashed}
\usepackage{mathtools}
\usepackage{amsfonts}
\usepackage{tikz}
\usetikzlibrary{arrows.meta,calc,angles,quotes}
\usepackage{xcolor}
\usepackage{physics}
\usepackage{comment}
\usepackage{subcaption}
\usepackage[percent]{overpic}
\usepackage{textgreek}
\usepackage{cite}
\usepackage{physics}
\usepackage{moresize}
\usepackage{dsfont}
\usepackage{dsdshorthand}
\usepackage{bbm}
\usepackage{booktabs}

\definecolor{darkgreen}{rgb}{0,0.5,0}
\definecolor{darkblue}{rgb}{0,0,0.6}
\definecolor{purple}{rgb}{0.4,.2,0.7}

\usepackage[colorlinks=true,citecolor=darkgreen,linkcolor=black,urlcolor=purple]{hyperref}

\usepackage{pdfsync}

\makeatletter
\newcommand*{\defeq}{\mathrel{\rlap{%
                     \raisebox{0.3ex}{$\m@th\cdot$}}%
                     \raisebox{-0.3ex}{$\m@th\cdot$}}%
                     =} 
\makeatother

\def\be{\begin{eqnarray}}
\def\ee{\end{eqnarray}}

\newcommand{\bea}{\begin{eqnarray}}
\newcommand{\eea}{\end{eqnarray}}
\def\ben{\begin{equation}}
\def\een{\end{equation}}

     \let\r=v

\def\be{\begin{equation}}
\def\ee{\end{equation}}
\def\ba{\begin{eqnarray}}
\def\ea{\end{eqnarray}}

\def\bal#1\eal{\begin{align}#1\end{align}}
\def\bs#1\es{\begin{split}#1\end{split}}

\allowdisplaybreaks  

\numberwithin{equation}{section}

\def\bx{\bar{X}}

\def\be{\begin{equation}}
\def\ee{\end{equation}}
\def\ba{\begin{eqnarray}}
\def\ea{\end{eqnarray}}
\def\bal#1\eal{\begin{align}#1\end{align}}

\def\r{\rightarrow}

\def\r{\right}

\def\lr{\leftrightarrow}

\def\br{\rbrace}

\def\ie{\begin{equation}\begin{aligned}}
\def\fe{\end{aligned}\end{equation}}

\usepackage{multirow}

\def \be {\begin{equation}}
\def \ee {\end{equation}}

\usepackage{framed}

\begin{document}

\onehalfspacing

\begin{center}

~
\vskip5mm

{\LARGE {Positivity constraints on the dynamics of \\cusped impurities in CFTs}
}

\vskip7mm
Jeevan Chandra

\vskip5mm

{\textit{Leinweber Institute for Theoretical Physics and Department of Physics, \\ University of California, Berkeley, USA
}
}

\vskip5mm

{\tt jcn1998@berkeley.edu}

\end{center}

\vspace{2mm}

\begin{abstract}

The cusp anomalous dimension is the universal logarithmic contribution to the free energy of a line defect with a cusp, and provides a useful observable characterizing defect dynamics. We generalize this notion to cusped line defects in the presence of a planar conformal boundary or interface. The geometry is characterized by three angles, and the distance of the cusp from the boundary or interface which separates short-distance and long-distance regimes. The resulting free energy contains two large logarithmic terms: a short-distance contribution governed by the usual vacuum cusp anomalous dimension, and a new long-distance contribution sensitive to the boundary or interface. Using reflection positivity for transverse deformations of the defect rays, we derive several constraints on the coefficients of these logarithmic terms, which can be packaged into negative semi-definiteness of their angular Hessians. We also derive constraints on the continuation of these coefficients to Lorentzian kinematics. We compute these coefficients and verify the constraints explicitly in free scalar and Maxwell theories and in planar $\mathcal N=4$ super-Yang-Mills theory at weak coupling for various choices of boundary conditions. At strong coupling, we compute the logarithmic coefficients in the holographic D3-D5 defect CFT and find a Gross-Ooguri-like phase transition between competing string worldsheets driven by the geometry of the cusp relative to the interface. We also observe a close connection of the cusp free energy in this setup with holographic corner entanglement entropy.

\end{abstract}

\pagebreak

\tableofcontents

\vspace{0.7in}

\section{Introduction and summary of results}
\label{sec:Intro}

When two conformal line-defect rays meet at a cusp with a specified opening angle, the defect free energy acquires a universal logarithmic contribution whose angle-dependent coefficient is the cusp anomalous dimension. It first arose in the study of renormalization of Wilson loops with cusps \cite{Polyakov:1980ca,Korchemsky:1987wg}, and has since become a useful observable because a single function encodes several kinds of defect data. Radial quantization identifies the cusp anomalous dimension with the appropriately normalized ground-state energy of two impurities on $S^{d-1}$ \cite{Cuomo:2024psk}. Near the smooth-line limit, its expansion is controlled by displacement-operator data and, for Wilson lines, by the Bremsstrahlung function \cite{Correa:2012at}; in the opposite sharp-cusp limit, it probes defect fusion and the associated Casimir energy \cite{Cuomo:2024psk,Diatlyk:2024qpr,Kravchuk:2024qoh}. Closely related cusp observables for replica twist defects control universal corner contributions to R\'enyi and entanglement entropies \cite{Casini:2006hu,Bianchi:2015liz}. In gauge theories, Lorentzian and lightlike cusps govern Sudakov logarithms and infrared factorization \cite{Korchemsky:1987wg,Korchemsky:1992xv}. The cusp anomalous dimension has been studied particularly extensively for Wilson lines in $\mathcal{N}=4$ SYM, from perturbation theory \cite{Makeenko:2006ds,Grozin:2015kna} and semiclassical strings at strong coupling \cite{Drukker:1999zq,Drukker:2011za} to exact approaches based on localization and integrability \cite{Correa:2012at,Correa:2012hh,Gromov:2015dfa}.

In this work, we generalize the cusp anomalous dimension by placing the cusp in the presence of a planar conformal boundary or interface. The cusp is located a perpendicular distance $y$ from the codimension-one locus. Using rotational invariance, the geometry is characterized by three angles as illustrated in figure \ref{fig:nonplanarcusp}: the elevation angles $\theta_L$ and $\theta_R$ of the two rays relative to the boundary/interface, and the azimuthal angle $\psi$ between the planes containing the two rays and the normal to the boundary/interface. The distance $y$ introduces a physical scale separating two asymptotic regimes. Denoting the UV cutoff by $a$ and the IR length of the defect by $L$, the universal large-logarithm part of the free energy takes the form
\begin{equation} \label{eq:intro-free-energy}
F(\theta_L,\theta_R,\psi;y)
=
\Gamma(\chi)\log\frac{y}{a}
+
A_{\rm IR}(\theta_L,\theta_R,\psi)\log\frac{L}{y}
+
O(1),
\end{equation}
where $\chi$ is the ordinary geometric opening angle between the two defect rays expressible in terms of the three angles (\ref{eq:chiangle}). At distances much shorter than $y$, the boundary or interface is invisible, and the coefficient of the UV logarithm is therefore the usual vacuum cusp anomalous dimension $\Gamma(\chi)$. At distances much larger than $y$, the defect probes the boundary/interface, and the coefficient $A_{\rm IR}$ contains new universal information about the long-distance interaction of the cusp with the boundary or interface. It is this second coefficient that will be our main observable.

In this paper, we derive a set of non-perturbative constraints on
$A_{\rm IR}$ following from Euclidean reflection positivity
\cite{OsterwalderSchrader1973}.
The basic ingredients are the displacement operators localized on the
two defect rays, whose correlation functions govern the response of
the defect free energy to transverse deformations of its embedding.
For a planar cusp whose geometry is illustrated in fig \ref{fig:planarcusp}, varying the two rays within the plane of the cusp
and applying reflection positivity to the corresponding integrated
displacement operators directly implies mixed concavity with respect
to the two ray angles. Out-of-plane deformations contain additional
information. Already in the absence of a boundary, in $d\geq 3$, they give a direct reflection-positivity proof of the monotonicity
of the vacuum cusp anomalous dimension, $\Gamma'(\chi)\geq0$. This
provides an alternative to the argument of \cite{Cuomo:2024psk},
where monotonicity follows from concavity together with the endpoint
condition at the smooth cusp. In the presence of a boundary or
interface, allowing both in-plane and out-of-plane deformations
organizes the positivity conditions into the statement that the mixed
angular Hessian of the free energy is negative semi-definite on the
reflection-symmetric locus. In rotationally invariant setups this
implies, in addition to concavity in the elevation angle, concavity and monotonicity in the azimuthal angle and a Cauchy-Schwarz-type inequality. Since the
two logarithmic scales in \eqref{eq:intro-free-energy} can be varied
independently, these conditions apply separately to their
coefficients. The vacuum cusp anomalous dimension already satisfies
the required constraints, so the genuinely new restrictions act on
the boundary/interface coefficient:
\begin{equation} \label{eq:introEucconstraints}
\partial_\psi A_{\rm IR}\geq 0,
\qquad
\partial_\psi^2 A_{\rm IR}\leq 0,
\qquad
\partial_{\theta_L}\partial_{\theta_R}A_{\rm IR}\leq 0,
\qquad
\bigl(\partial_{\theta_L}\partial_{\theta_R}A_{\rm IR}\bigr)
\bigl(\partial_\psi^2 A_{\rm IR}\bigr)
-
\left|\partial_{\theta_L}\partial_\psi A_{\rm IR}\right|^2
\geq 0,
\end{equation}
evaluated on the reflection-symmetric slice $\theta_L=\theta_R$. Beyond these inequalities, our explicit examples mentioned below exhibit evidence for a stronger infinite hierarchy in the azimuthal angle. On the reflection-symmetric locus, we find that the negatives of the logarithmic coefficients with nontrivial $\psi$-dependence obey an alternating-sign derivative hierarchy,
\begin{equation}
(-1)^{n+1}\partial_\psi^n \Gamma(\chi)\geq 0, \qquad (-1)^{n+1}\partial_\psi^n A_{\rm IR}\geq 0\qquad n\geq 1.
\end{equation}
We do not presently have a general derivation of this property and therefore regard it as a conjecture. The first two members of the hierarchy are precisely the monotonicity and concavity conditions that follow from reflection positivity, whereas the higher-derivative inequalities do not follow from our two-point argument.

We also obtain a Lorentzian constraint by analytically continuing the azimuthal angle into a rapidity parameter as $\psi\rightarrow\pi-i\rho$ while keeping the elevation angles fixed. Geometrically, this turns the two rays into spacelike defects lying in opposite Rindler wedges while the boundary/interface becomes a timelike hypersurface invariant under Rindler reflection. Rindler positivity is the Lorentzian analog of Euclidean reflection positivity: for operators supported in one Rindler wedge, the matrix of correlators with their anti-linear Rindler-reflected images is positive semi-definite. We assume that this property extends to the defect-ray operators considered here (see \cite{Cuomo:2026mop} for a discussion of this extension). The resulting Cauchy–Schwarz inequality gives, for the symmetric cusp and assuming the Lorentzian continuation is real,
\begin{equation}
A_{\rm IR}^{\,L}(\theta,\theta,\rho)
\geq
A_{\rm IR}^{\,L}(\theta,\theta,\rho=0)=A_{\rm IR}(\theta,\theta,\psi=\pi).
\label{eq:intro-lorentzian-constraint}
\end{equation}
This is the boundary/interface analogue of the recently derived positivity constraint on the spacelike Lorentzian vacuum cusp anomalous dimension \cite{Cuomo:2026mop}. Similarly, if we assume that Osterwalder–Schrader reflection positivity applies directly to the Euclidean defect-ray insertions, the corresponding Cauchy–Schwarz inequality yields a finite-angle bound on \(A_{\rm IR}\),
\begin{equation}
    2A_{\rm IR}(\theta_L, \theta_R, \pi-\psi_L-\psi_R) \geq A_{\rm IR}(\theta_L,\theta_L,\pi-2\psi_L) + A_{\rm IR}(\theta_R, \theta_R, \pi-2\psi_R)\,.
\end{equation}
Expanding about the reflection-symmetric locus reproduces (\ref{eq:introEucconstraints}), while for finitely separated ray configurations it gives additional constraints beyond these local Hessian inequalities.

We compute $A_{\rm IR}$ in a collection of free and weakly coupled examples: $4d$ scalar and Maxwell theories and planar $\mathcal{N}=4$ SYM, and verify both the Euclidean and Lorentzian constraints explicitly for various conformal boundary conditions. Finally, we compute $A_{\rm IR}$ in the strongly coupled D3-D5 defect CFT \cite{KarchRandall2001,DeWolfeFreedmanOoguri2002}, where the cusped Maldacena-Wilson line is described by a classical fundamental-string worldsheet \cite{Maldacena1998WilsonLoops,Rey:1998ik}. The worldsheet ending on the probe D5-brane yields an angular generalization of the interface-particle potential \cite{NagasakiTanidaYamaguchi2012}. We further extend this holographic calculation to the generalized
Maldacena-Wilson cusp, allowing the scalar polarization of each defect
ray to be tilted relative to the $S^2$ wrapped by the D5-brane. This
yields an interface-dependent strong-coupling generalization of the
usual geometric/internal-angle cusp \cite{Drukker:2011za}, described by
a D5-ending worldsheet with a non-trivial profile in $S^5$. Furthermore, there is a competing saddle described by a worldsheet that does not end on the D5-brane that evaluates to the vacuum cusp anomalous dimension, and the two saddles exchange dominance as the cusp angles are varied. This gives a new Gross-Ooguri-like phase transition, analogous to the original connected/disconnected worldsheet transition \cite{GrossOoguri1998,Zarembo1999WilsonLoopCorrelator}, but driven here by the geometry of a single cusp relative to an interface. Related worldsheet transitions have appeared for antiparallel Wilson lines in the D3-D5 defect CFT \cite{PretiTrancanelliVescovi2017} and for circular Wilson loops in the same theory \cite{BonanseaDavoliGriguoloSeminara2020}. We also find that the D5-ending solution is closely related to the
holographic corner function for entanglement entropy in
AdS$_4$/BCFT$_3$ \cite{Seminara:2017hhh}. Under this correspondence,
the Gross-Ooguri-like transition of the cusped Wilson line maps to
the holographic entanglement transition in which connected and
disconnected Ryu-Takayanagi surfaces exchange dominance as the wedge
geometry is varied.

\section{Positivity constraints on the cusp free energy} \label{sec:positivityconstraints}

In this section, we derive general non-perturbative constraints on the free energy of a cusped defect in the presence of a boundary using reflection positivity. We begin with the planar configuration, for which reflection positivity implies a mixed concavity condition with respect to the two angles made by the defect rays with the boundary. We then consider the general non-planar cusp and show that the mixed Hessian of the free energy with respect to deformations of the two rays is negative semi-definite on the reflection-symmetric locus. In rotationally invariant setups, this translates into a pair of concavity conditions, a Cauchy--Schwarz-type inequality, and a monotonicity condition for the dependence on the azimuthal angle between the two defect planes. Finally, for scale-invariant configurations, we separate the free energy into UV and IR logarithmic contributions and show that these constraints apply independently to their coefficients. The UV contribution is governed by the ordinary boundaryless cusp anomalous dimension and automatically satisfies the required inequalities, so the genuinely new constraints act on the IR contribution describing the interaction of the cusp with the boundary.

Although we use the language of a ``boundary'' throughout the derivation, the same reflection-positivity argument applies when the codimension-one boundary is replaced by a conformal interface. Indeed, the Osterwalder--Schrader reflection used below acts by reflecting one defect ray into the other while leaving the codimension-one locus invariant; the argument does not rely on spacetime terminating at that locus. Thus, provided the defect CFT is unitary and reflection positive, the same constraints apply to the free energy of a cusped defect in the presence of an interface.

\subsection{Concavity from in-plane deformations} \label{sec:planarcusp}

\begin{figure}
    \centering

    \begin{tikzpicture}[scale=0.8,
    transform shape,
        x=1cm,
        y=1cm,
        line cap=round,
        line join=round,
        >=Latex
    ]

    \def\rayL{5.3}
    \def\rayR{5.3}
    \def\thone{32}
    \def\thtwo{32}
    \def\ysep{3.3}
    \def\xdash{6.3}
    \def\xbnd{6.4}
    \def\angrad{1.35}

    \coordinate (P) at (0,0);
    \coordinate (L) at
        ({-\rayL*cos(\thone)},{\rayL*sin(\thone)});
    \coordinate (R) at
        ({ \rayR*cos(\thtwo)},{\rayR*sin(\thtwo)});
    \coordinate (B1) at (-\xbnd,-\ysep);
    \coordinate (B2) at ( \xbnd,-\ysep);

    \draw[dashed,line width=1pt]
        (-\xdash,0) -- (\xdash,0);

    \draw[line width=1.3pt]
        (L) -- (P) -- (R);

    \draw[-Latex,line width=1.2pt]
        (P) -- (0,-\ysep);

    \node[right]
        at ($(P)!0.52!(0,-\ysep)$)
        {\Large $y$};

    \draw[line width=1.3pt]
        (B1) -- (B2);

    \draw[line width=1pt]
        ({\angrad*cos(180-\thone)},
         {\angrad*sin(180-\thone)})
        arc[
            start angle=180-\thone,
            end angle=180,
            radius=\angrad
        ];

    \node at (-2.10,0.8)
        {\Large $\theta_L$};

    \draw[line width=1pt]
        (\angrad,0)
        arc[
            start angle=0,
            end angle=\thtwo,
            radius=\angrad
        ];

    \node at (2.10,0.8)
        {\Large $\theta_R$};

    \end{tikzpicture}

    \caption{
        The geometry of a planar cusped defect in the presence of a
        boundary. The cusp is located a perpendicular distance $y$
        from the boundary, and the two defect rays make angles
        $\theta_L$ and $\theta_R$ with the direction parallel to the
        boundary.
    }
    \label{fig:planarcusp}
\end{figure}
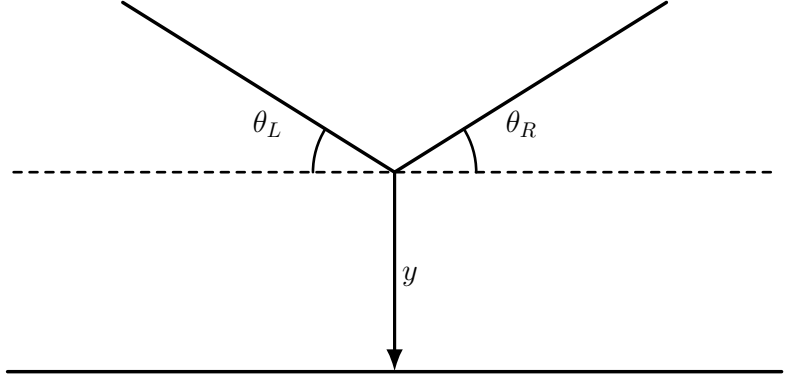

In this subsection, we will analyze the setup described in figure \ref{fig:planarcusp} of a cusped defect in the presence of a boundary and show that the free energy obeys a non-perturbative mixed concavity condition using reflection positivity. Here, we are assuming that the plane of the cusp contains the normal to the boundary. But, more generally, the plane of the cusp need not contain the normal. We will analyze this general case in the next subsection.

As shown in figure \ref{fig:planarcusp}, the setup involves a d-dimensional unitary Euclidean BCFT on a half-space\footnote{This setup of a cusped defect in the presence of a conformal boundary was introduced in \cite{Abdalla:2026wdx} for 2d CFTs.},
\begin{equation}
\mathbb H^d=\{(x,\vec x_\parallel,x_\perp)\in \mathbb R^d \,:\, x_\perp\ge 0\},
\end{equation}
with conformal boundary at $x_\perp=0$. $\vec x_\parallel$ are the remaining coordinates parallel to the boundary.
The cusp is located at the point
\begin{equation}
P=(0,\vec 0,y), \qquad y>0,
\end{equation}
and the two defect rays lie in the $(x,x_\perp)$-plane. A convenient parametrization is
\begin{align}
X_L(s)&=P+s\,(-\cos\theta_L,\sin\theta_L), \qquad s\ge 0, \\
X_R(s)&=P+s\,(\cos\theta_R,\sin\theta_R), \qquad s\ge 0.
\end{align}
Here $\theta_L$ and $\theta_R$ are the angles made by the two rays with the conformal boundary. We denote the line defect operator by $\mathcal{D}(\theta_L,\theta_R;y)$. Its free energy is
\begin{equation}
F(\theta_L,\theta_R;y)\equiv -\log \langle \mathcal D(\theta_L,\theta_R;y)\rangle.
\end{equation}
When the two angles are equal, $\theta_L=\theta_R$,
the configuration is invariant under the reflection $R: x\mapsto -x$,
which preserves the half-space, preserves the conformal boundary, and exchanges the two defect rays.

To relate angle derivatives of the free energy to local operators on the defect, we use the displacement operator. Under an infinitesimal deformation of the defect embedding,
\begin{equation}
X^\mu(s)\to X^\mu(s)+\delta X^\mu(s),
\end{equation}
the response is governed by the displacement operator $D_\mu(s)$ localized on the defect. For the angular deformations considered here, only the normal component enters. Let
\begin{equation}
\widehat D_i(s)\equiv n_i^\mu(s)\,D_\mu(s),
\end{equation}
where $n_i^\mu$ is the unit normal to the $i$th arm in the $(x,x_\perp)$-plane, chosen so that an increase in $\theta_i$ corresponds to the deformation
\begin{equation} 
\delta X_i^\mu(s)=s\,\delta\theta_i\, n_i^\mu(s).
\end{equation}
Then the first variation of the defect free energy takes the form
\begin{equation}
\partial_{\theta_i}F(\theta_L,\theta_R;y)
=
-\int_0^L ds\, s\,\langle \widehat D_i(s)\rangle.
\end{equation}
Differentiating once more, we obtain
\begin{equation} \label{eq:mixed-second-derivative-displacement}
\partial_{\theta_L}\partial_{\theta_R}F(\theta_L,\theta_R;y)
=
-\int_0^L ds_1\int_0^L ds_2\,
s_1 s_2\,
\langle \widehat D_L(s_1)\widehat D_R(s_2)\rangle_c.
\end{equation}
Since the insertions lie on different rays, there are no same-arm contact terms in this mixed derivative.

We now specialize to the reflection-symmetric locus $\theta_L=\theta_R$. Define the integrated displacement operators along the defect rays,
\begin{equation} \label{eq:intdisp}
\mathcal O_R\equiv \int_0^L ds\, s\,\widehat D_R(s),
\qquad
\mathcal O_L\equiv \int_0^L ds\, s\,\widehat D_L(s).
\end{equation}
By construction, the reflection $R$ exchanges the two rays and therefore maps $\Theta(\mathcal O_R)=\mathcal O_L$ where $\Theta$ is the reflection operator that implements the geometric reflection $R$ along with the Hermitian-conjugation of the 
Equation \eqref{eq:mixed-second-derivative-displacement} can then be rewritten as
\begin{equation}
\partial_{\theta_L}\partial_{\theta_R}F(\theta_L,\theta_R;y)\big|_{\theta_L=\theta_R}
=
-\langle \mathcal O_L \mathcal O_R\rangle_c.
\end{equation}
Now we apply reflection positivity. For any operator $\mathcal O$ supported in the region $x>0$, Osterwalder-Schrader positivity \cite{OsterwalderSchrader1973} implies
\begin{equation}
\langle \Theta(\mathcal O)\,\mathcal O\rangle \ge 0,
\end{equation}
Since we are interested in the connected 2-point function, we apply this to the fluctuation $\widetilde{\mathcal O}_R\equiv \mathcal O_R-\langle \mathcal O_R\rangle$,
and obtain
\begin{equation} 
\langle \Theta(\widetilde{\mathcal O}_R)\,\widetilde{\mathcal O}_R\rangle \ge 0 \implies \langle \mathcal O_L \mathcal O_R\rangle_c\ge 0.
\end{equation}
Therefore,
\begin{equation} \label{eq:mixed-concavity-free-energy}
\partial_{\theta_L}\partial_{\theta_R}F(\theta_L,\theta_R;y)\Big|_{\theta_L=\theta_R}
\le 0.
\end{equation}
This is the mixed concavity condition.

\subsection{General constraints from out-of-plane deformations} \label{sec:nonplanarcusp}

We can derive more general constraints on the free energy by imposing reflection positivity on out-of-plane deformations of the defect rays. As a warm-up, let us first consider the boundaryless case where we show that the monotonicity of the cusp anomalous dimension follows from reflection positivity of out-of-plane deformations thereby providing a novel proof of monotonicity. Let the plane formed by the two defect rays meeting at the cusp be spanned by units vectors $\Hat{v}_1$ and $\Hat{v}_2$ where $\Hat{v}_1$ is along the angular bisector between the two rays and $\Hat{v}_2$ is orthogonal to it. The tangent vectors along the defect rays can then be expressed in this basis as
\begin{equation}
\begin{split}
    t_L(\alpha)= & \cos\left(\frac{\alpha}{2}\right) \Hat{v}_1+\sin\left(\frac{\alpha}{2}\right) \Hat{v}_2 \\
    t_R(\alpha)= & \cos\left(\frac{\alpha}{2}\right) \Hat{v}_1-\sin\left(\frac{\alpha}{2}\right) \Hat{v}_2
    \end{split} \,.
\end{equation}
Here, $L,R$ are the labels for the two defect rays and $\alpha$ is the opening angle of the cusp. Let $\Hat{v}_3$ be an out-of-plane direction in which we can deform the two defect rays. If we deform the defect rays by angles $\beta_{L,R}$ respectively along $\Hat{v}_3$, then the new tangent vectors along the rays are 
\begin{equation}
    \begin{split}
        \Tilde{t}_L(\alpha, \beta_L)= & \cos(\beta_L)t_L(\alpha)+\sin(\beta_L) \Hat{v}_3\\
        \Tilde{t}_R(\alpha, \beta_R)= & \cos(\beta_R)t_R(\alpha)+\sin(\beta_R) \Hat{v}_3
    \end{split}.
\end{equation}
Reflection positivity implies that 
\begin{equation} \label{eq:reflectionpositivityoutplane}
    \frac{\partial^2 F}{\partial \beta_L \partial \beta_R}\bigg |_{\beta_L=\beta_R}\leq 0.
\end{equation}
So far, we have not assumed rotational symmetry of the CFT. Now, if we impose the  SO$(d)$ rotational symmetry, the free energy should only depend on the angle between the tangent vectors,
\begin{equation}
    \Tilde{t}_L. \Tilde{t}_R\equiv \cos(\Tilde{\alpha})= \cos(\beta_L)\cos(\beta_R)\cos(\alpha)+\sin(\beta_L)\sin(\beta_R)\,.
\end{equation}
Evaluating (\ref{eq:reflectionpositivityoutplane}) at $\beta_L=\beta_R=0$ gives the monotonicity condition,
\begin{equation}
    -\frac{\Gamma'(\alpha
    )}{\sin(\alpha)}\leq 0 \implies \Gamma'(\alpha)\geq 0\,.
\end{equation}
In \cite{Cuomo:2024psk}, the above condition was derived using $\Gamma''(\alpha)\leq 0$ and the end-point condition $\Gamma'(\pi)=0$.

\begin{figure}
    \centering

    \begin{tikzpicture}[
        scale=0.85,
        transform shape,
        line cap=round,
        line join=round,
        >=Latex,
        every node/.style={font=\Large},
        every pic quotes/.style={font=\Large}
    ]

    \coordinate (P) at (0,0);

    \coordinate (A) at (-6.8,-1.15);
    \coordinate (B) at ( 6.8,-1.15);
    \coordinate (C) at ( 5.3, 0.10);
    \coordinate (D) at (-5.3, 0.10);

    \coordinate (E) at (-6.4,-3.10);
    \coordinate (F) at ( 6.4,-3.10);
    \coordinate (G) at ( 5.0,-2.10);
    \coordinate (H) at (-5.0,-2.10);

    \coordinate (mLend) at (-5.1,-0.55);
    \coordinate (mRend) at ( 5.1,-0.55);

    \coordinate (QL) at (-4.10,-0.43);
    \coordinate (QR) at ( 4.10,-0.43);

    \coordinate (TL) at (-4.10,4.85);
    \coordinate (TR) at ( 4.10,4.85);

    \coordinate (N) at (0,6.20);

    \fill[gray!15] (E) -- (F) -- (G) -- (H) -- cycle;
    \draw[line width=1pt] (E) -- (F) -- (G) -- (H) -- cycle;

    \draw[densely dotted,line width=1pt] (A) -- (B) -- (C) -- (D) -- cycle;

    \draw[densely dotted,line width=1pt] (TL) -- (TR);
    \draw[densely dotted,line width=1pt] (TL) -- (QL);
    \draw[densely dotted,line width=1pt] (TR) -- (QR);

    \draw[-Latex,line width=1.4pt] (P) -- (N);
    \node[right] at ($(N)+(0,-0.25)$) {$\hat n$};

    \draw[-Latex,line width=1.4pt,black] (P) -- (TL);
    \draw[-Latex,line width=1.4pt,black]  (P) -- (TR);

    \node[black] at (-4.45,5.15) {$t_L$};
    \node[black]  at ( 4.45,5.15) {$t_R$};

    \draw[-Latex,dashed,line width=1.2pt,black] (P) -- (mLend);
    \draw[-Latex,dashed,line width=1.2pt,black]  (P) -- (mRend);

    \node[black] at (-5.55,-0.45) {$\hat m_L$};
    \node[black]  at ( 5.55,-0.45) {$\hat m_R$};

    \fill (P) circle (2.2pt);
    \node[below] at (P) {$P$};


    \def\rtheta{2.10}

    \draw[-Latex,line width=1pt]
        ($(P)+(186.15:\rtheta)$)
        arc[
            start angle=186.15,
            end angle=130.22,
            radius=\rtheta
        ];

    \node at ($(P)+(158.2:2.55)$)
        {$\theta_L$};

    \draw[-Latex,line width=1pt]
        ($(P)+(-6.15:\rtheta)$)
        arc[
            start angle=-6.15,
            end angle=49.78,
            radius=\rtheta
        ];

    \node at ($(P)+(21.8:2.55)$)
        {$\theta_R$};

    \coordinate (psiL) at ($(P)!0.47!(mLend)$);
    \coordinate (psiR) at ($(P)!0.47!(mRend)$);

    \draw[Latex-Latex,line width=1pt]
        (psiL)
        .. controls (-1.75,-0.72) and (1.75,-0.72) ..
        (psiR);

    \node at (0,-0.82)
        {$\psi$};

    \end{tikzpicture}

    \caption{
       The figure describes the geometry of the non-planar cusp. The defect rays meet at the cusp denoted P. $\theta_{L,R}$ are the elevation angles of the defect rays in their respective defect planes. $\psi$ is the azimuthal angle between the defect planes. The cusp P is located at a distance $y$ above the shaded boundary plane.
    }
    \label{fig:nonplanarcusp}
\end{figure}
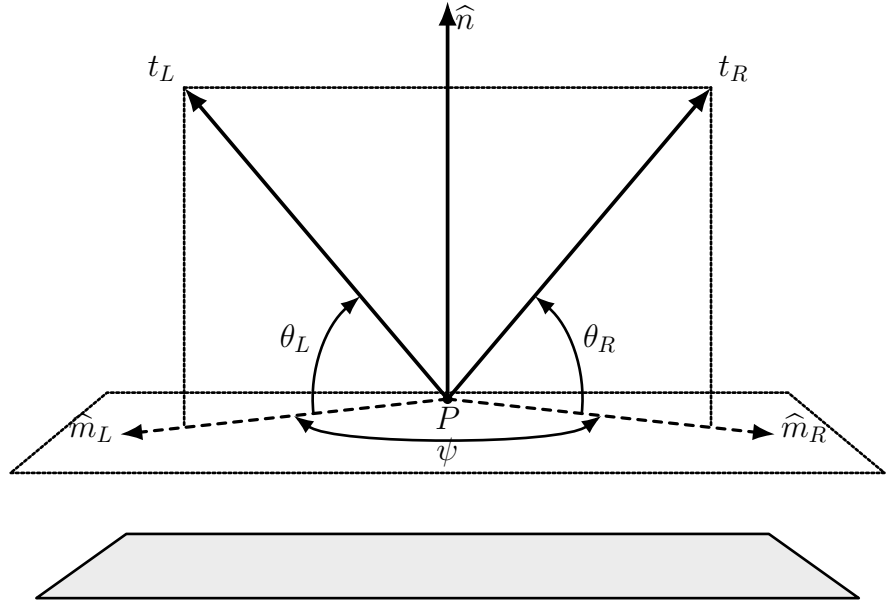

Now, let us go to the boundary case. First, let us describe the geometry of the setup in detail. See figure \ref{fig:nonplanarcusp} for illustration. Let $\Hat{n}$ be inward-pointing normal vector to the boundary and let $(\Hat{e}_1, \Hat{e}_2)$ be basis vectors on the boundary. Each of the defect rays forms a plane with the normal vector $\Hat{n}$. Let the planes intersect the boundary along rays $\Hat{m}_{L,R}$ respectively making angles of $\psi_{L,R}$ with $\Hat{e}_1$,
\begin{equation}
    \begin{split}
        \Hat{m}_L= & -\cos(\psi_L)\Hat{e}_1-\sin(\psi_L)\Hat{e}_2 \\
        \Hat{m}_R= & \cos(\psi_R)\Hat{e}_1-\sin(\psi_R)\Hat{e}_2
    \end{split}
\end{equation}
Rotational symmetry of the boundary would mean that the free energy depends on the two azimuthal angles $\psi_{L,R}$ only in terms of the angle between the two defect planes $\psi\equiv \pi-(\psi_L+\psi_R)$. On these planes, the defect rays are elevated at angles $\theta_{L,R}$ respectively so that the tangent vectors along the rays can be expressed as
\begin{equation}
    \begin{split}
        & t_L= \cos(\theta_L) \Hat{m}_L+\sin(\theta_L) \Hat{n} \\
        & t_R=\cos(\theta_R) \Hat{m}_R+ \sin(\theta_R) \Hat{n}\,.
    \end{split}
\end{equation}
We parametrise the angles as a pair of 2-vectors associated with each defect ray,
\begin{equation} \label{eq:anglevectors}
    \lambda_L=(\theta_L,\psi_L), \qquad \lambda_R=(\theta_R,\psi_R)\,.
\end{equation}
In the space of these 4 angles, the reflection symmetric locus is described by
\begin{equation}
    \lambda_L=\lambda_R \implies \theta_L=\theta_R, \quad \psi_L=\psi_R\,.
\end{equation}
Reflection positivity imposes the following negative semi-definiteness condition on the Hessian of the free energy treated as a $2\times 2$ matrix on this locus,
\begin{equation}
    \frac{\partial^2 F}{\partial \lambda_L^a \partial \lambda_R^b}\bigg |_{\lambda_L=\lambda_R}= -\langle \Theta(\mathcal{O}^a_R)\mathcal{O}^b_R\rangle_c \implies \frac{\partial^2 F}{\partial \lambda_L \partial \lambda_R}\bigg |_{\lambda_L=\lambda_R}\preceq 0\,.
\end{equation}
Here, $\Theta$ is the reflection operator that implements the reflection of the integrated displacement operators in (\ref{eq:intdisp}).
In components, this implies a pair of concavity conditions and a Cauchy-Schwarz type inequality coming from the requirement that the determinant of the above Hessian matrix is non-negative,
\begin{equation}
\begin{split}
    & \frac{\partial^2 F}{\partial \theta_L \partial \theta_R}\bigg |_{\lambda_L=\lambda_R}\leq 0, \qquad  \frac{\partial^2 F}{\partial \psi_L \partial \psi_R}\bigg |_{\lambda_L=\lambda_R}\leq 0\\
    & \frac{\partial^2 F}{\partial \theta_L \partial \theta_R}\bigg |_{\lambda_L=\lambda_R}\frac{\partial^2 F}{\partial \psi_L \partial \psi_R}\bigg |_{\lambda_L=\lambda_R}-\left|\frac{\partial^2 F}{\partial \theta_L \partial \psi_R}\right|^2_{\lambda_L=\lambda_R} \geq 0\,.
    \end{split}
\end{equation}
In rotationally symmetric settings, the free energy only depends on three angles $\theta_L,\theta_R,\psi$, so the above conditions can be written as
\begin{equation}
    \begin{split}
    & \frac{\partial^2 F}{\partial \theta_L \partial \theta_R}\bigg |_{\theta_L=\theta_R}\leq 0, \qquad  \frac{\partial^2 F}{\partial \psi^2}\bigg |_{\theta_L=\theta_R}\leq 0\\
    & \frac{\partial^2 F}{\partial \theta_L \partial \theta_R}\bigg |_{\theta_L=\theta_R}\frac{\partial^2 F}{\partial \psi^2}\bigg |_{\theta_L=\theta_R}-\left|\frac{\partial^2 F}{\partial \theta_L \partial \psi}\right|^2_{\theta_L=\theta_R} \geq 0\,.
    \end{split}\,.
\end{equation}
These exhaust the constraints arising from reflection positivity in $3d$. In $d\geq 4$, there are further $d-3$ transverse directions parallel to the boundary that we could deform the defect rays into. Reflection positivity for those transverse deformations give additional constraints just like in the boundary-less case, namely negative-semidefiniteness of the transverse Hessian of the free energy,
\begin{equation} \label{eq:refpostransbound}
     \frac{\partial^2 F}{\partial \beta_L \partial \beta_R}\bigg |_{\beta_L=\beta_R}\leq 0\,.
\end{equation}
Here, $\beta_{L,R}$ are the $(d-3)$-component vectors of transverse angles for each defect ray. We will now determine how the condition (\ref{eq:refpostransbound}) constrains the free energy when there is SO$(d-3)$ rotational symmetry in these transverse directions. We will follow the same steps as we did to derive the monotonicity for the boundaryless case. Let $\Hat{e}_3$ be a transverse direction that we can deform the defect rays into, then the new tangent vectors along the rays are 
\begin{equation}
    \begin{split}
        \Tilde{t}_L= & \cos(\beta_L)t_L+\sin(\beta_L) \Hat{e}_3\\
        \Tilde{t}_R= & \cos(\beta_R)t_R+\sin(\beta_R) \Hat{e}_3
    \end{split}.
\end{equation}
The projections of these new tangent vectors onto the plane parallel to the boundary are
\begin{equation}
    \begin{split}
        \Tilde{m}_L=  \Tilde{t}_L-( \Tilde{t}_L.\Hat{n})\Hat{n}= & \cos(\beta_L)\cos(\theta_L)\Hat{m}_L+\sin(\beta_L) \Hat{e}_3\\
        \Tilde{m}_R= \Tilde{t}_R-( \Tilde{t}_R.\Hat{n})\Hat{n}= & \cos(\beta_R) \cos(\theta_R)\Hat{m}_R+\sin(\beta_R) \Hat{e}_3
    \end{split}.
\end{equation}
The deformed physical angles on which the free energy depends are denoted $(\Tilde{\theta}_L,\Tilde{\theta}_R, \Tilde{\psi})$ and are given by
\begin{equation}
\begin{split}
   & \Tilde{t}_L.\Hat{n}\equiv \sin(\Tilde{\theta}_L)=\cos(\beta_L)\sin(\theta_L)\\
   & \Tilde{t}_R.\Hat{n}\equiv \sin(\Tilde{\theta}_R)=\cos(\beta_R)\sin(\theta_R)\\
   & \frac{\Tilde{m}_L.\Tilde{m}_R}{|\Tilde{m}_L|\,|\Tilde{m}_R|}\equiv \cos(\Tilde{\psi})=\frac{\cos(\beta_L)\cos(\beta_R)\cos(\theta_L)\cos(\theta_R)\cos(\psi)+\sin(\beta_L)\sin(\beta_R)}{\sqrt{\sin^2(\beta_L)+\cos^2(\beta_L)\cos^2(\theta_L)}\sqrt{\sin^2(\beta_R)+\cos^2(\beta_R)\cos^2(\theta_R)}}\,.
    \end{split}
\end{equation}
Now, we evaluate the constraint from reflection positivity on the undeformed reflection-symmetric slice,
\begin{equation}
    \frac{\partial^2 F}{\partial \beta_L\partial \beta_R}\bigg |_{\beta_L=\beta_R=0, \theta_L=\theta_R}\leq 0\,.
\end{equation}
Since $F=F(\Tilde{\theta}_L,\Tilde{\theta}_R,\Tilde{\psi})$, note that the angular derivatives on the undeformed slice are given by
\begin{equation}
    \begin{split}
        &\frac{\partial \Tilde{\theta}_L}{\partial \beta_L}\bigg |_{\beta_L=\beta_R=0}= \frac{\partial \Tilde{\theta}_R}{\partial \beta_R}\bigg |_{\beta_L=\beta_R=0}=0,\\
        & \frac{\partial^2 \Tilde{\psi}}{\partial \beta_L\partial \beta_R}\bigg |_{\beta_L=\beta_R=0}=-\frac{1}{\cos(\theta_L)\cos(\theta_R)\sin(\psi)}\,,
    \end{split}
\end{equation}
so only the derivatives of the azimuthal angle contribute. Thus, we have
\begin{equation}
     \frac{\partial^2 F}{\partial \beta_L\partial \beta_R}\bigg|_{\beta_L=\beta_R=0, \theta_L=\theta_R}=-\frac{1}{\cos^2(\theta)\sin(\psi)}\frac{\partial F}{\partial \psi}\bigg|_{\theta_L=\theta_R} \leq 0\,. 
\end{equation}
Therefore, we have derived the monotonicity condition with respect to the azimuthal angle,
\begin{equation}
    \frac{\partial F}{\partial \psi}\bigg|_{\theta_L=\theta_R}\geq 0\,.
\end{equation}
Although this proof of monotonicity only works in $d\geq 4$, the above monotonicity condition continues to hold also in $d=3$ because of concavity in $\psi$ and assuming differentiability at the end-point,
\begin{equation}
    \frac{\partial^2 F}{\partial \psi^2}\leq 0, \quad \frac{\partial F(\theta_L,\theta_R,\psi=\pi)}{\partial \psi}=0 \implies \frac{\partial F}{\partial \psi}\geq 0\,.
\end{equation}
This exhausts all the constraints on the free energy from reflection positivity of the two-point function of the displacement operator in any dimension.

\subsection{Constraints on large logarithms in the free energy} \label{sec:Freeenergyscale}

So far, we have expressed the constraints in terms of the free energy. In this section, we will use the fact that the only length scale is the distance of the cusp to the boundary to argue that the free energy is dominated by two large logarithms and thereby translate the constraints on the free energy into constraints on the two logarithmic coefficients. Recall that the second derivative of the free energy can be written in terms of the integrated two-point function of displacement operator,
\begin{equation}
    \frac{\partial^2 F}{\partial \lambda_L^a \partial \lambda_R^b}=-\int_a^L ds s\int_a^L dt t \langle D_a(s)D_b(t) \rangle_c\,.
\end{equation}
The displacement operator for a conformal line defect has scaling dimension $2$. Since the only length scale is the distance of the cusp to the boundary $y$ (apart from the UV and IR cutoffs), we can rescale the distance variables in the integration by $y$ thereby introducing dimensionless variables, $s=y e^{\sigma}$ and $t=y e^{\tau}$. Now, the only dependence on $y$ is through the integration limits,
\begin{equation}
     \frac{\partial^2 F}{\partial \lambda_L^a \partial \lambda_R^b}=-\int_{-u}^v d\tau \int_{-u}^v d\sigma K_{ab}(\tau, \sigma), \qquad u\equiv\log\frac{y}{a}, \quad v\equiv\log\frac{L}{y}\,,
\end{equation}
where $K_{ab}$ is the dimensionless kernel constructed from the two-point function of the displacement operator. In the two asymptotic limits when we are probing the cusp at very short and very long distances, we assume that the kernel approaches a translation-invariant form,
\begin{equation}
    K_{ab}(\tau, \sigma) \to \begin{cases}
        & K^{\rm UV}_{ab}(\tau-\sigma), \qquad \tau,\sigma \ll 0 \\
        & K_{ab}^{\rm IR}(\tau-\sigma), \qquad \tau, \sigma \gg 0\,.
    \end{cases}
\end{equation}
Therefore, it is suggestive to split the double-integral over the kernel in a way that it is dominated by the two asymptotic limits,
\begin{equation}
\begin{split}
    \int_{-u}^v d\tau \int_{-u}^v d\sigma K_{ab}(\tau, \sigma)= & \int_{-u}^{O(1)}d\tau d\sigma K^{\rm UV}_{ab}(\tau-\sigma) + \int_{O(1)}^{v}d\tau d\sigma K^{\rm IR}_{ab}(\tau-\sigma)+\dots \\
    =& \kappa^{\rm UV}_{ab}u+\kappa^{\rm IR}_{ab} v+\dots \,.
    \end{split}
\end{equation}
The $\dots$ denote the terms coming from the crossover regions where one or both of $\sigma, \tau$ are $O(1)$. Since we are taking the UV cutoff $a \to 0$ and the IR cutoff $L \to \infty$, we expect the crossover contributions are negligible in comparison to the leading logarithms. In addition, we assume clustering of the connected two-point function of the displacement operator, which excludes quadratic or higher powers of the logarithms, as in the standard vacuum cusp analysis of \cite{Cuomo:2024psk}. With these assumptions, we can express the renormalized free energy (after removing possible power-law divergence terms in the cutoffs) as a sum of two large logarithms (upto O(1) terms) with angle dependent coefficients that govern the dynamics of the cusp in the presence of a boundary,
\begin{equation} \label{eq:largelogdecomp}
    F= A_{\rm UV}\log \left(\frac{y}{a} \right) + A_{\rm IR} \log \left(\frac{L}{y} \right)+O(1)\,.
\end{equation}
Since the boundary is invisible to the UV coefficient, it is given by the same function $\Gamma(\chi)$ that governs the cusp anomalous dimension in the boundaryless case. Here, $\chi$ is the angle between the tangent vectors along the defect rays
\begin{equation} \label{eq:chiangle}    \cos(\chi)=\cos(\theta_L)\cos(\theta_R)\cos(\psi)+\sin(\theta_L)\sin(\theta_R)\,.
\end{equation}
Therefore, the free energy can be expressed as
\begin{equation}
     F(\theta_L,\theta_R,\psi;y)= \Gamma(\chi)\log \left(\frac{y}{a} \right) + A_{\rm IR}(\theta_L,\theta_R,\psi) \log \left(\frac{L}{y} \right)+O(1)\,.
\end{equation}
Since the positivity constraints we derived are linear in the free energy, and since the two logarithms can be tuned independently, the constraints should apply independently to both the coefficients. We will first show that concavity and monotonicity of $\Gamma$, namely $\Gamma''(\chi)\leq 0$ and $\Gamma'(\chi)\geq 0$ automatically satisfy the new constraints. To this end, the monotonicity in $\psi$ follows from
\begin{equation}
    \frac{\partial \Gamma}{\partial \psi} = \Gamma'(\chi)\frac{\cos^2(\theta)\sin(\psi)}{\sin(\chi)}\geq 0\,.
\end{equation}
Next, to see the negative semi-definiteness of the Hessian of $\Gamma$, there is a very useful way of writing the Hessian on the reflection-symmetric slice $\theta_L=\theta_R \equiv \theta$,
\begin{equation}
    H \equiv \begin{bmatrix}
        & \frac{\partial^2 \Gamma}{\partial \theta_L \partial \theta_R} & \frac{\partial^2 \Gamma}{\partial \theta_L \partial \psi} \\
        & \frac{\partial^2 \Gamma}{\partial \theta_R \partial \psi} & \frac{\partial^2 \Gamma}{\partial \psi^2}
    \end{bmatrix}=\Gamma''(\chi)u u^T-\Gamma'(\chi) w w^T
\end{equation}
where the vectors $u$ and $w$ are given by
\begin{equation}
    u=\frac{1}{\sin(\chi)}\begin{bmatrix}
        -\sin(\theta)\cos(\theta)(1-\cos(\psi))\\
        \cos^2(\theta) \sin(\psi)
    \end{bmatrix}, \qquad w=\frac{1}{(\sin(\chi))^{\frac{3}{2}}}\begin{bmatrix}
        \cos(\theta)\sin(\psi) \\
        \sin(\theta)\cos^2(\theta)(1-\cos(\psi))
    \end{bmatrix}
\end{equation}
From the above representation, the negativity of the diagonal entries of the Hessian is manifest and note that the determinant can be expressed as
\begin{equation}
    \det H= -\Gamma'(\chi)\Gamma''(\chi)(u_1w_2-u_2w_1)^2 \geq 0\,.
\end{equation}
Hence, we have verified that $H \preceq 0$. This shows that positivity imposes new constraints only on the IR coefficient $A_{\rm IR}$,
\begin{equation} \label{eq:IRpositivity}
    \partial_\psi A_{\rm IR}\geq 0, \quad \partial^2_\psi A_{\rm IR}\leq 0, \quad \partial_{\theta_L}\partial_{\theta_R}A_{\rm IR}\leq 0, \quad (\partial_{\theta_L}\partial_{\theta_R}A_{\rm IR})(\partial^2_\psi A_{\rm IR})-|\partial_{\theta_L}\partial_\psi A_{\rm IR}|^2 \geq 0.
\end{equation}
All the inequalities above are to be evaluated on the reflection-symmetric slice $\theta_L=\theta_R$.

\subsection{Bounds on Lorentzian continuations of logarithmic coefficients}

So far, we have derived constraints on the logarithmic coefficients in the free energy using reflection positivity of Euclidean cusped defect correlators in the presence of a boundary. In this section, we will analytically continue the configuration to Lorentzian signature such that the defect rays meeting at the cusp are spacelike separated and the boundary is a timelike hypersurface. Placing the defect rays in different Rindler wedges, we will assume Rindler positivity of the defect correlator to derive a bound on the Lorentzian continuation of the logarithmic coefficients in the free energy. This will be a generalization of the result in the recent paper \cite{Cuomo:2026mop} where they assume Rindler positivity to derive that the Lorentzian continuation of the vacuum cusp anomalous dimension $\Gamma^L(\rho)$ is non-negative. $\Gamma^L(\rho)$ is analytically continued from the Euclidean result $\Gamma(\theta)$ using $\theta\to \pi-i\rho$ where $\rho$ is the rapidity. In our setup, we will argue that it is natural to analytically continue the azimuthal angle $\psi \to \pi -i\rho$ to derive analogous bounds.

First, we will describe the geometry of the setup in Lorentzian signature. Only three of the spacetime coordinates $(t,x,x_\perp, \dots)$ will be important for the discussion so we will drop the remaining transverse coordinates. In these coordinates, the cusp is placed at $P=(0,0,y)$ with the timelike boundary at $x_\perp=0$. A defect ray in the right Rindler wedge $x>|t|$ at an elevation of angle $\theta$ relative to the boundary and with rapidity $\rho$ is parametrised by
\begin{equation}
    X_R(s)= P+ s\left(\cos\theta\sinh\rho, \cos\theta\cosh\rho, \sin\theta\right), \qquad s\geq 0\,.
\end{equation}
A defect ray in the left Rindler wedge can be parametrised by Rindler reflection $(t,x,x_\perp)\to (-t,-x,x_\perp)$ of a defect ray in the right Rindler wedge. Note importantly that Rindler reflection leaves the boundary invariant.
The Minkowski inner product between the tangent vector of a defect ray in the right Rindler wedge with that of a defect ray in the left Rindler wedge is therefore given in terms of the elevation angles $\theta_{L,R}$ and the rapidities $\rho_{L,R}$ by
\begin{equation}
    \Dot{X}_R.\Dot{X}_L = -\cos(\theta_L)\cos(\theta_R)\cosh(\rho_L-\rho_R)+\sin(\theta_L)\sin(\theta_R)\,.
\end{equation}
Comparing with the angle between the defect rays in the Euclidean configuration given in (\ref{eq:chiangle}) suggests the following analytic continuation of the azimuthal angle to rapidity,
\begin{equation}
    \psi \to \pi-i(\rho_R-\rho_L)\,.
\end{equation}

Recall that Rindler positivity is the Lorentzian analogue of Euclidean reflection positivity. Let $\Theta_R$ denote the anti-linear operation that combines the geometric Rindler reflection
\begin{equation}
(t,x,x_\perp)\mapsto(-t,-x,x_\perp)
\end{equation}
with Hermitian conjugation. For any finite collection of operators $\mathcal O_i$ supported in the right Rindler wedge, Rindler positivity states that the following matrix of two-point functions is positive semi-definite,
\begin{equation}
\langle \Theta_R(\mathcal O_i)\mathcal O_j\rangle \succeq 0\,.
\end{equation}
We assume that this property extends to the defect-ray operators in the present setup. Since the Rindler reflection leaves the timelike boundary invariant, its application to the defect rays gives\footnote{For Wightman functions of local operators in Lorentzian QFT, Rindler positivity has been established in \cite{Casini:2010bf}; see also the foundational Bisognano–Wichmann results \cite{BisognanoWichmann1975,BisognanoWichmann1976}. We assume that Rindler positivity extends to the defect-ray operators considered here; \cite{Cuomo:2026mop} provides a justification for defects admitting suitable endpoint operators, while a general proof for arbitrary defect lines is not presently available.},
\begin{equation} \label{eq:refpos}
    M_{ij} \equiv \langle \Theta_R \left[\mathcal{D}(\theta_i,\rho_i;y)\right]\mathcal{D}(\theta_j,\rho_j;y)\rangle_{\rm BCFT} \succeq 0\,.
\end{equation}
Here, the indices $i,j$ are the ray labels in the right Rindler wedge. $\mathcal{D}$ is the defect operator associated with a ray ending at the cusp. $\Theta_R$ is the Rindler-conjugation operator acting on the defect operator that performs the geometric Rindler reflection and the Hermitian conjugation. The Cauchy-Schwarz inequality from (\ref{eq:refpos}) bounds the Lorentzian partition function $Z_L$ by the Euclidean partition function for the planar cusp $(\psi=\pi$),
\begin{equation} \label{eq:CauchySchwarz}
    |Z_L(\theta_L,\theta_R, \rho)|^2\leq Z(\theta_L,\theta_L,\pi)Z(\theta_R,\theta_R,\pi)\,,
\end{equation}
where the Lorentzian partition function is defined by the analytic continuation of the Euclidean partition function,
\begin{equation}
    Z_L(\theta_L,\theta_R, \rho)\equiv Z(\theta_L,\theta_R,\pi-i\rho)\,.
\end{equation}
Specializing to the symmetric cusp $\theta_L=\theta_R$, Cauchy-Schwarz (\ref{eq:CauchySchwarz}) translates into the following inequalities between the Lorentzian and Euclidean logarithmic coefficients of the free energy,
\begin{equation} \label{eq:Lorconstraints}
   \text{Re}\, A^L_{\rm UV}(\theta, \theta, \rho) \geq A^L_{\rm UV}(\theta, \theta, \rho=0)= A_{\rm UV}(\theta, \theta, \pi), \qquad \text{Re}\, A^L_{\rm IR}(\theta, \theta, \rho) \geq A^L_{\rm IR}(\theta, \theta, \rho=0)= A_{\rm IR}(\theta, \theta, \pi)\,.
\end{equation}
So, the logarithmic coefficients for the Lorentzian spacelike cusp are bounded from below by the corresponding logarithmic coefficients for the Euclidean planar cusp. In general, the Lorentzian continuations of the logarithmic coefficients need not be real but whenever the Lorentzian partition function obeys Hermitian analyticity, $Z_L(\theta, \theta, \rho)^*=Z_L(\theta, \theta, -\rho)$, and is invariant under $\rho\to -\rho$, the logarithmic coefficients are real. This will be the case in all the examples discussed in the next section. Also, it is important to note that (\ref{eq:Lorconstraints}) imposes no new constraints on the Lorentzian continuation of the vacuum cusp anomalous dimension $\Gamma^L$ given its non-negativity already derived in \cite{Cuomo:2026mop}. So, the only novel constraint in (\ref{eq:Lorconstraints}) is the one on the Lorentzian continuation of the IR coefficient.

Note that, if we analogously assume that Osterwalder–Schrader reflection positivity applies directly to the Euclidean defect-ray operators, then the resulting Cauchy-Schwarz inequality takes the form,
\begin{equation} \label{eq:CSpart}
    |Z(\lambda_L,\lambda_R)|^2 \leq Z(\lambda_L,\lambda_L)Z(\lambda_R,\lambda_R)\,,
\end{equation}
where $\lambda_L$ and $\lambda_R$ are the angle 2-vectors associated with each defect ray introduced in (\ref{eq:anglevectors}). When the corresponding free energy is separated into the two large logarithms, it gives the following constraint on the IR coefficient,
\begin{equation} \label{eq:CSAIR}
    2A_{\rm IR}(\theta_L, \theta_R, \pi-\psi_L-\psi_R) \geq A_{\rm IR}(\theta_L,\theta_L,\pi-2\psi_L) + A_{\rm IR}(\theta_R, \theta_R, \pi-2\psi_R)\,.
\end{equation}
Expanding the above inequality about the reflection-symmetric locus $\lambda_L=\lambda_R$ reproduces the negative semi-definiteness constraints on the angular Hessian derived in Sec \ref{sec:nonplanarcusp}. For finite separations away from this locus, however, (\ref{eq:CSAIR}) provides additional constraints. It is important to emphasize that (\ref{eq:CSAIR}) assumes that Osterwalder–Schrader reflection positivity extends directly to the nonlocal defect-ray insertions, which has not been established in general, unlike the constraints of Sec \ref{sec:Freeenergyscale} derived from reflection positivity of local displacement operators. Note also that the corresponding inequality on the UV coefficient i.e., the vacuum cusp anomalous dimension follows from the usual concavity and monotonicity constraints and hence provides no new constraints.

\section{The cusp free energy in some free or weakly-coupled examples} \label{sec:examples}

In this section, we compute the UV and IR logarithmic coefficients in the cusp free energy and verify the positivity constraints derived in Sec \ref{sec:positivityconstraints} in some examples of conformal line defects in free or weakly-coupled $4d$ field theories: a free scalar, a free Maxwell field, and planar $\mathcal{N}=4$ SYM at weak coupling. In these simple examples, the cusp free energy can be computed easily using the method of images, allowing us to directly check the mixed concavity condition and, for the general non-planar cusp, the negative semi-definiteness and monotonicity constraints derived in Sec \ref{sec:positivityconstraints}. Furthermore, we analytically continue the expressions for the logarithmic coefficients to Lorentzian kinematics and verify the Lorentzian bounds (\ref{eq:Lorconstraints}) as well.

\subsection{Free massless scalar field in $4d$} \label{sec:scalar}

In this subsection, we will first compute the logarithmic coefficients of the free energy for the simple example of a cusped pinning defect in a free massless scalar field in $4d$ with a conformal boundary/interface using the method of images. Then, we will show that the expressions for these coefficients satisfy all the positivity constraints derived in Sec \ref{sec:positivityconstraints}. 

Consider a free massless scalar on the half-space
\begin{equation}
\mathbb H^4=\{(x_1,x_2,x_3,z)\,:\, z\ge 0\},
\end{equation}
with action
\begin{equation}
S=\frac12\int_{z\ge 0} d^4x\, (\partial \phi)^2 .
\end{equation}
We impose either Neumann or Dirichlet boundary conditions at $z=0$:
\begin{equation}
\partial_z \phi|_{z=0}=0 \qquad (\text{Neumann}), 
\qquad\qquad
\phi|_{z=0}=0 \qquad (\text{Dirichlet}) .
\end{equation}
It is convenient to write the propagator uniformly as
\begin{equation} \label{eq:scalarpropagator}
G_\eta(x,x')=\frac{1}{4\pi^2}\left(\frac{1}{|x-x'|^2}+\eta\,\frac{1}{|x-x'^*|^2}\right),
\qquad
\eta=
\begin{cases}
+1 & \text{Neumann},\\
-1 & \text{Dirichlet},
\end{cases}
\end{equation}
where
\begin{equation}
x'^*=(x_1',x_2',x_3',-z')
\end{equation}
is the mirror point.

We now introduce the pinning defect
\begin{equation}
\mathcal D_h[\Sigma]=\exp\!\left[-h\int_\Sigma ds\, \phi(X(s))\right].
\end{equation}
In $d=4$ the scalar has dimension $\Delta_\phi=1$, so $h$ is dimensionless. The cusp is placed at
\begin{equation}
P=(0,0,0,y), \qquad y>0,
\end{equation}
and the two defect rays are parameterized by
\begin{equation}
    \begin{split}
        X_1(s)&=(-s\cos\theta_1,\,0,\,0,\,y+s\sin\theta_1), \qquad s\in[0,L],\\
X_2(t)&=(\phantom{-}t\cos\theta_2,\,0,\,0,\,y+t\sin\theta_2), \qquad t\in[0,L].
    \end{split}
\end{equation}
The free energy is given in terms of the propagator (\ref{eq:scalarpropagator}) by
\begin{equation}
F_\eta(\theta_1,\theta_2;L,y)\equiv -\log \langle \mathcal D_h[\Sigma]\rangle_\eta=
-\frac{h^2}{2}\int_\Sigma ds \int_\Sigma ds'\, G_\eta(X(s),X(s'))\,.
\end{equation}

To compute the free energy and verify the positivity constraints, it is useful to separate the various pairings of the two rays and their images. Writing
\begin{equation}
F_\eta=
-\frac{h^2}{2}\Big(J_{11}+J_{22}+2J_{12}+\eta(K_{11}+K_{22}+2K_{12})\Big),
\end{equation}
the direct contributions are
\begin{equation}
\begin{split}
J_{11}=J_{22}
&=\frac{1}{4\pi^2}\int_0^L ds \int_0^L dt\, \frac{1}{(s-t)^2},\\
J_{12}
&=\frac{1}{4\pi^2}\int_0^L ds \int_0^L dt\, \frac{1}{s^2+t^2+2st\cos(\theta_1+\theta_2)} ,
\end{split}
\end{equation}
while the image contributions are
\begin{equation}
\begin{split}
K_{11}
&=\frac{1}{4\pi^2}\int_0^L ds \int_0^L dt\,
\frac{1}{s^2+t^2-2st\cos 2\theta_1+4y(s+t)\sin\theta_1+4y^2},\\
K_{22}
&=\frac{1}{4\pi^2}\int_0^L ds \int_0^L dt\,
\frac{1}{s^2+t^2-2st\cos 2\theta_2+4y(s+t)\sin\theta_2+4y^2},\\
K_{12}
&=\frac{1}{4\pi^2}\int_0^L ds \int_0^L dt\,
\frac{1}{s^2+t^2+2st\cos(\theta_1-\theta_2)+4y(s\sin\theta_1+t\sin\theta_2)+4y^2}.
\end{split}
\end{equation}
The same-ray terms $J_{11},J_{22}$ require a short-distance cutoff $a$ and give
\begin{equation}
J_{11}=J_{22}
=
\frac{1}{2\pi^2}\left(\frac{L}{a}-1-\log\frac{L}{a}\right).
\end{equation}
The direct interaction between the two rays has a logarithmic divergence and yields
\begin{equation}
J_{12}
=
\frac{1}{4\pi^2}\,\frac{\theta_1+\theta_2}{\sin(\theta_1+\theta_2)}\,
\log\frac{L}{a}+O(1).
\end{equation}
The image terms are finite in the UV because the boundary separation $y$ cuts off the short-distance singularity. Their logarithmic large-$L$ parts are
\begin{equation}
\begin{split}
K_{11}
&=
\frac{1}{4\pi^2}\,\frac{\pi-2\theta_1}{\sin 2\theta_1}\,\log\frac{L}{y}+O(1),\\
K_{22}
&=
\frac{1}{4\pi^2}\,\frac{\pi-2\theta_2}{\sin 2\theta_2}\,\log\frac{L}{y}+O(1),\\
K_{12}
&=
\frac{1}{4\pi^2}\,\frac{\theta_1-\theta_2}{\sin(\theta_1-\theta_2)}\,\log\frac{L}{y}+O(1)\,.
\end{split}
\end{equation}
Collecting the logarithmic and linearly divergent pieces, we obtain
\begin{align} \label{eq:freeenergyscalar}
F_\eta
&=
-\frac{h^2}{2\pi^2}\left(\frac{L}{a}-1-\log\frac{L}{a}\right)
-\frac{h^2}{4\pi^2}\,\frac{\theta_1+\theta_2}{\sin(\theta_1+\theta_2)}\,
\log\frac{L}{a}
\nonumber\\
&\qquad
-\frac{\eta h^2}{8\pi^2}
\left[
\frac{\pi-2\theta_1}{\sin 2\theta_1}
+\frac{\pi-2\theta_2}{\sin 2\theta_2}
+2\,\frac{\theta_1-\theta_2}{\sin(\theta_1-\theta_2)}
\right]\log\frac{L}{y},
\end{align}
with $\eta=+1$ for Neumann and $\eta=-1$ for Dirichlet.

To isolate the angular interaction of the cusp, it is convenient to subtract the free energies of the two isolated rays in the same half-space. This removes $J_{11},J_{22},K_{11},K_{22}$ and leaves
\begin{equation}
F_{\eta,\rm int}
=
-\frac{h^2}{4\pi^2}\left[\,\frac{\theta_1+\theta_2}{\sin(\theta_1+\theta_2)}\,
\log\frac{L}{a}
+\eta\,\frac{\theta_1-\theta_2}{\sin(\theta_1-\theta_2)}\,
\log\frac{L}{y}\right].
\end{equation}
Differentiating twice, we see that
\begin{equation}
\partial_{\theta_1}\partial_{\theta_2}F_{\eta,\rm int}\Big|_{\theta_1=\theta_2}\le 0
\end{equation}
for both Neumann and Dirichlet boundary conditions, in agreement with the mixed concavity condition derived from reflection positivity.

The above calculation was for the planar cusp. But we can readily generalize it to compute the free energy for the general cusp configuration which is a function of three angles $(\theta_1,\theta_2,\psi)$. The interaction part of the free energy can be expressed as
\begin{equation}
    F_{\eta,\rm int}
=
-\frac{h^2}{4\pi^2}\int_{0}^L ds \int_0^L dt \left[\frac{1}{|s\Hat{v}_1-t\Hat{v}_2|^2}+\eta \frac{1}{|2y \Hat{n}+s\Hat{v}_1-t R\Hat{v}_2|^2}\right]
\end{equation}
The first term is the direct interaction between the defect rays and the second term is the interaction between a defect ray and the image of the other defect ray. The vectors $\Hat{v}_{1,2}$ are respectively the unit tangent vectors along the defect rays pointed away from the cusp, and $\Hat{n}$ is the inward-pointing unit normal to the boundary. $R$ is the reflection operator acting on these vectors. The inner products relevant for the above computation are
\begin{equation} \label{eq:directimageangles}
    \begin{split}
        & \Hat{v}_1.\Hat{v}_2\equiv \cos \chi_d= \cos \theta_1 \cos \theta_2 \cos\psi+\sin \theta_1 \sin \theta_2 \\
        & \Hat{v}_1. R \Hat{v}_2\equiv \cos \chi_i= \cos \theta_1 \cos \theta_2 \cos\psi-\sin \theta_1 \sin \theta_2\,.
    \end{split}
\end{equation}
The above inner products define the direct and image angles in terms of which we can express the interaction free energy compactly as
\begin{equation} \label{eq:intfreenergyscalar}
     F_{\eta,\rm int}
=
-\frac{h^2}{4\pi^2}\left[g(\chi_d) \log \frac{y}{a}+(g(\chi_d)+\eta g(\chi_i))\log \frac{L}{y}\right], \qquad  g(x)\equiv \frac{\pi-x}{\sin x}\,.
\end{equation}
The UV coefficient of the interaction free energy is the non-factorizing part of the vacuum cusp anomalous dimension; restoring the subtracted factorizing contribution gives the standard vacuum cusp anomalous dimension $\Gamma(\chi_d)$. The logarithmic coefficient which characterizes the non-factorizing interaction of the cusp with the boundary is given by
\begin{equation}
    A_{\rm IR, int}=-\frac{h^2}{4\pi^2}\left(g(\chi_d)+\eta g(\chi_i)\right)\,.
\end{equation}
Note that the complete expression for $A_{\rm IR}$ will also involve the factorizing terms subtracted from (\ref{eq:freeenergyscalar}).
The UV coefficient is given by the vacuum cusp anomalous dimension $\Gamma(\chi_d)$ as expected.
We checked numerically that the Hessian for each coefficient is negative semi-definite for both Dirichlet and Neumann boundary conditions. In fact, we observed numerically that the free energy in this setup obeys a stronger condition: the higher derivatives of either logarithmic coefficient with respect to the azimuthal angle $\psi$ evaluated on the reflection-symmetric slice $\theta_1=\theta_2$ alternate in sign.
Equivalently, we can say that $- \Gamma(\chi_d)$ and $A_{\rm IR}$ are completely monotonic functions of the azimuthal angle,
\begin{equation}
    (-1)^{n+1} \frac{\partial^n  \Gamma(\chi_d)}{\partial \psi^n}\bigg |_{\theta_1=\theta_2}\geq 0, \qquad  (-1)^{n+1} \frac{\partial^n  A_{\rm IR}}{\partial \psi^n}\bigg |_{\theta_1=\theta_2}\geq 0\,,
\end{equation}
as illustrated in table \ref{tab:scalar-complete-monotonicity} for some chosen values of the parameters.
Note that we have also checked numerically that the negative of the vacuum cusp anomalous dimension is completely monotonic in the opening angle $\chi_d$.\footnote{See also \cite{Henn:2024positivity} for discussions of complete monotonicity in other cusp variables.} But the observation of complete monotonicity of $-\Gamma(\chi_d)$ in $\psi$ is a strictly stronger result that does not follow from complete monotonicity in $\chi_d$ given the map (\ref{eq:directimageangles}) between $\chi_d$ and $\psi$. Motivated by this observation and the analogous results in the other examples in this section, we conjecture that this alternating-sign hierarchy has a more general origin. In addition to verifying the above inequalities along the reflection-symmetric locus, we also numerically verified that the inequality (\ref{eq:CSAIR}) is satisfied for various angles away from the reflection-symmetric locus for both the Dirichlet and Neumann boundary conditions.

\begin{table}
\centering
\scriptsize
\setlength{\tabcolsep}{4pt}

\textbf{(a) $\Gamma(\chi)$ : derivatives in the opening angle $\chi$}

\vspace{2mm}

\resizebox{\textwidth}{!}{%
\begin{tabular}{c c c c c c c c c}
\hline
$(\theta,\psi)$ & $\chi$
& $C_1$ & $C_2$ & $C_3$ & $C_4$ & $C_5$ & $C_6$ & $C_7$ \\
\hline
$(0.20,\,0.50)$ & $0.490$
& $12.7$ & $53.3$ & $327$ & $2.67{\times}10^{3}$
& $2.73{\times}10^{4}$ & $3.34{\times}10^{5}$ & $4.78{\times}10^{6}$ \\

$(0.25,\,2.70)$ & $2.48$
& $0.246$ & $0.449$ & $0.392$ & $0.871$
& $1.49$ & $4.11$ & $10.8$ \\

$(0.60,\,1.10)$ & $0.892$
& $3.62$ & $8.72$ & $29.6$ & $133$
& $748$ & $5.03{\times}10^{3}$ & $3.95{\times}10^{4}$ \\

$(0.90,\,2.20)$ & $1.17$
& $1.98$ & $3.79$ & $9.80$ & $33.7$
& $144$ & $735$ & $4.38{\times}10^{3}$ \\

$(1.25,\,1.50)$ & $0.433$
& $16.3$ & $77.1$ & $535$ & $4.94{\times}10^{3}$
& $5.70{\times}10^{4}$ & $7.89{\times}10^{5}$ & $1.28{\times}10^{7}$ \\
\hline
\end{tabular}%
}

\vspace{4mm}

\textbf{(b) $\Gamma(\chi)$: derivatives in the azimuthal angle $\psi$}

\vspace{2mm}

\resizebox{\textwidth}{!}{%
\begin{tabular}{c c c c c c c c}
\hline
$(\theta,\psi)$
& $C_1$ & $C_2$ & $C_3$ & $C_4$ & $C_5$ & $C_6$ & $C_7$ \\
\hline
$(0.20,\,0.50)$
& $12.4$ & $51.1$ & $308$ & $2.46{\times}10^{3}$
& $2.46{\times}10^{4}$ & $2.95{\times}10^{5}$ & $4.14{\times}10^{6}$ \\

$(0.25,\,2.70)$
& $0.160$ & $0.396$ & $0.237$ & $0.647$
& $0.811$ & $2.52$ & $5.17$ \\

$(0.60,\,1.10)$
& $2.82$ & $5.64$ & $15.5$ & $56.7$
& $258$ & $1.41{\times}10^{3}$ & $8.95{\times}10^{3}$ \\

$(0.90,\,2.20)$
& $0.670$ & $1.02$ & $1.17$ & $2.44$
& $5.21$ & $14.8$ & $46.1$ \\

$(1.25,\,1.50)$
& $3.86$ & $6.00$ & $11.7$ & $31.6$
& $105$ & $420$ & $1.96{\times}10^{3}$ \\
\hline
\end{tabular}%
}

\vspace{4mm}

\textbf{(c) $A_{\rm IR}$: Dirichlet boundary condition}

\vspace{2mm}

\resizebox{\textwidth}{!}{%
\begin{tabular}{c c c c c c c c}
\hline
$(\theta,\psi)$
& $C_1$ & $C_2$ & $C_3$ & $C_4$ & $C_5$ & $C_6$ & $C_7$ \\
\hline
$(0.20,\,0.50)$
& $6.74$ & $41.2$ & $305$ & $2.66{\times}10^{3}$
& $2.69{\times}10^{4}$ & $3.13{\times}10^{5}$ & $4.22{\times}10^{6}$ \\

$(0.25,\,2.70)$
& $0.0161$ & $0.0429$ & $0.0477$ & $0.145$
& $0.280$ & $0.974$ & $2.68$ \\

$(0.60,\,1.10)$
& $2.22$ & $5.38$ & $16.1$ & $59.0$
& $263$ & $1.41{\times}10^{3}$ & $8.88{\times}10^{3}$ \\

$(0.90,\,2.20)$
& $0.551$ & $0.898$ & $1.20$ & $2.54$
& $5.41$ & $15.0$ & $46.0$ \\

$(1.25,\,1.50)$
& $3.82$ & $6.00$ & $11.7$ & $31.6$
& $105$ & $420$ & $1.96{\times}10^{3}$ \\
\hline
\end{tabular}%
}

\vspace{4mm}

\textbf{(d) $A_{\rm IR}$: Neumann boundary condition}

\vspace{2mm}

\resizebox{\textwidth}{!}{%
\begin{tabular}{c c c c c c c c}
\hline
$(\theta,\psi)$
& $C_1$ & $C_2$ & $C_3$ & $C_4$ & $C_5$ & $C_6$ & $C_7$ \\
\hline
$(0.20,\,0.50)$
& $18.1$ & $61.0$ & $310$ & $2.26{\times}10^{3}$
& $2.24{\times}10^{4}$ & $2.78{\times}10^{5}$ & $4.05{\times}10^{6}$ \\

$(0.25,\,2.70)$
& $0.304$ & $0.749$ & $0.425$ & $1.15$
& $1.34$ & $4.06$ & $7.67$ \\

$(0.60,\,1.10)$
& $3.42$ & $5.91$ & $14.9$ & $54.4$
& $253$ & $1.41{\times}10^{3}$ & $9.02{\times}10^{3}$ \\

$(0.90,\,2.20)$
& $0.789$ & $1.14$ & $1.14$ & $2.35$
& $5.02$ & $14.6$ & $46.2$ \\

$(1.25,\,1.50)$
& $3.89$ & $6.00$ & $11.6$ & $31.6$
& $105$ & $420$ & $1.96{\times}10^{3}$ \\
\hline
\end{tabular}%
}

\caption{
The table illustrates the alternating-sign derivative
hierarchy for the cusp logarithms in the free scalar example. All angles are given in radians.
In (a), we define
$C_n\equiv(-1)^{n+1}\partial_\chi^n\Gamma(\chi)$,
with the opening angle $\chi=\chi(\theta,\psi)$ determined by
$\cos\chi=\cos^2\theta\cos\psi+\sin^2\theta$.
In (b), we instead define
$C_n\equiv(-1)^{n+1}\partial_\psi^n\Gamma(\chi(\theta,\psi))$
at fixed $\theta$.
In (c) and (d),
$C_n\equiv(-1)^{n+1}\partial_\psi^n A_{\rm IR}$
for Dirichlet and Neumann boundary conditions, respectively.
We have suppressed the common positive overall normalization factors,
which do not affect the signs of the derivatives. Notice that all the displayed signed derivatives are positive.
}
\label{tab:scalar-complete-monotonicity}
\end{table}

The above analysis can readily be generalized to a conformal interface placed along $z=0$ separating the two copies of the free scalar theories on $z>0$ and $z<0$ respectively,
\begin{equation}
S=\frac12\int_{z> 0} d^4x\, (\partial \phi)^2+\frac12\int_{z< 0} d^4x\, (\partial \phi)^2 .
\end{equation}
A one-parameter family of scale-invariant linear gluing conditions for the scalar field across the interface are given by \cite{HerzogSchaub2024},
\begin{equation}
    \phi_+(x,0)=\lambda\phi_-(x,0), \qquad \partial_z \phi_+(x,0)=\lambda^{-1}\partial_z \phi_-(x,0)\,,
\end{equation}
where $\lambda \in \mathbb{R}- \{0\}$ is a dimensionless parameter. We place the pinning defect on the $+$ side, so the relevant propagator for the scalar field in the presence of the interface is given by
\begin{equation}
    G_r(x,x')=\frac{1}{4\pi^2}\left(\frac{1}{|x-x'|^2}+r\frac{1}{|x-x^{'*}|^2}\right), \qquad r=\frac{\lambda^2-1}{\lambda^2+1}\,.
\end{equation}
The reflection coefficient $r$ is bounded, $r \in [-1,1]$. So, the interaction free energy of the cusp interpolates between the Dirichlet and Neumann free energy for the conformal boundary case,
\begin{equation}
F_{r,\rm int}
=
-\frac{h^2}{4\pi^2}\left[\,\frac{\theta_1+\theta_2}{\sin(\theta_1+\theta_2)}\,
\log\frac{L}{a}
+r\,\frac{\theta_1-\theta_2}{\sin(\theta_1-\theta_2)}\,
\log\frac{L}{y}\right].
\end{equation}
for the planar cusp and
\begin{equation}
     F_{r,\rm int}
=
-\frac{h^2}{4\pi^2}\left[g(\chi_d) \log \frac{y}{a}+(g(\chi_d)+r g(\chi_i))\log \frac{L}{y}\right], \qquad  g(x)\equiv \frac{\pi-x}{\sin x}\,.
\end{equation}
for the general non-planar cusp. So, the logarithmic coefficient that characterizes the non-factorizing interaction of the cusp with the interface is given by
\begin{equation} \label{eq:intAIRscalar}
    A_{\rm IR, int}=-\frac{h^2}{4\pi^2}\left(g(\chi_d)+rg(\chi_i)\right)\,.
\end{equation}
Since it interpolates between the Dirichlet and Neumann values which satisfy all the positivity constraints, it also satisfies all the positivity constraints due to convexity of the negative semi-definite Hessian matrices.

Now, we turn to verification of the bound (\ref{eq:Lorconstraints}) on the Lorentzian continuation of the logarithmic coefficient for the symmetric cusp with equal elevation angles $\theta_L=\theta_R\equiv \theta$. First, note that under the continuation of the azimuthal angle to rapidity $\psi\to \pi-i\rho$, the direct and image angles as given in (\ref{eq:directimageangles}) continue to 
\begin{equation} \label{eq:Lordirimgangles}
   \begin{split}
       & \cos \chi_d \to z_d(\theta,\rho)\equiv -\cos^2 \theta \cosh \rho + \sin^2\theta,\\
       & \cos \chi_i \to z_i(\theta, \rho) \equiv -\cos^2 \theta \cosh \rho - \sin^2\theta\,.
   \end{split}
\end{equation}
From the above expressions, we see that after the analytic continuation, $\cos \chi_d$ and $\cos \chi_i$ are no longer restricted to lie between $(-1,1)$. A suitable analytic continuation of the function $g(x)$ that is valued on this extended domain is given by the following integral representation,
\begin{equation} \label{eq:scalarfz}
    g(x) \to f(z) \equiv \int_0^{\infty} \frac{du}{\cosh u-z}, \quad z<1\,.
\end{equation}
As a consistency check, the above integral representation for $f(z)$ gives $g(x)$ when $z \in (-1,1)$ with $z=\cos x$. We therefore have the Lorentzian continuation of (\ref{eq:intAIRscalar}),
\begin{equation}
    A^L_{\rm IR, int}(\theta,\rho) = -\frac{h^2}{4\pi^2}(f(z_d)+rf(z_i))\,.
\end{equation}
Since $r\in [-1,1]$, we can check that the above function is monotonically increasing in $\rho$, 
\begin{equation}
    \frac{\partial  A^L_{\rm IR, int}(\theta,\rho) }{\partial \rho} \geq 0, \qquad \rho \geq 0\,.
\end{equation}
This means the bound (\ref{eq:Lorconstraints}) is satisfied since it can be expressed as
\begin{equation}
    A^L_{\rm IR, int}(\theta,\rho) - A^L_{\rm IR, int}(\theta,\rho=0) \geq 0\,.
\end{equation}
In fact, we can show that the above difference is bounded from above by the value at large rapidity $\rho \to \infty$,
\begin{equation}
    \frac{h^2}{4\pi^2}\left[\frac{2\theta}{\sin 2\theta}+r\right] \geq A^L_{\rm IR, int}(\theta,\rho) - A^L_{\rm IR, int}(\theta,\rho=0) \geq 0\,
\end{equation}
Note that adding back the terms in the free energy which factorize between the defect arms will leave the above inequality unchanged since they simply cancel between the LHS and the RHS. 

\subsection{Free Maxwell field in \texorpdfstring{$4d$}{4d}} \label{sec:Maxwell}

In this subsection, we will compute the logarithmic coefficients for a charged Wilson line with a cusp in free Maxwell theory in $4d$ for Dirichlet, Neumann and also for the duality-rotated boundary conditions in the presence of a non-zero $\vartheta$-angle. For these cases, we will show that all the positivity constraints are satisfied.

Consider the free Maxwell field on the half-space $\mathbb{H}^{4}$,
with Euclidean action
\begin{equation}
S
=
\int_{z\ge 0} d^4x\,
\left[
\frac{1}{4e^2}F_{\mu\nu}F^{\mu\nu}
\right],
\qquad
F_{\mu\nu}=\partial_\mu A_\nu-\partial_\nu A_\mu .
\end{equation}
Later in this subsection, we will also consider the effect of adding a $\vartheta$-term to the action. We denote the dual field strength as
\begin{equation}
\widetilde F_{\mu\nu}\equiv \frac12 \epsilon_{\mu\nu\rho\sigma}F^{\rho\sigma},
\qquad
\widetilde F_{za}=\frac12 \epsilon_{abc}F^{bc},
\qquad a,b,c=1,2,3 .
\end{equation}
Varying the action gives
\begin{equation}
\delta S\big|_{\partial}
=
-\int_{z=0} d^3x\,
\left[
\frac{1}{e^2}F_{za}
\right]\delta A_a .
\end{equation}
This leads to the following local conformal boundary conditions:
\begin{equation}
    \begin{split}
        &\text{Dirichlet:} \quad
\widetilde F_{za}\big|_{z=0}=0,
\\
&\text{Neumann:} \quad
F_{za}\big|_{z=0}=0,
    \end{split}
\end{equation}
We work in Feynman gauge. For Dirichlet and Neumann boundary conditions, the half-space
propagator is obtained by the method of images:
\begin{equation} \label{eq:maxwellprop}
\langle A_\mu(x)A_\nu(x')\rangle_{\eta}
=
\frac{e^2}{4\pi^2}
\left[
\frac{\delta_{\mu\nu}}{|x-x'|^2}
+
\eta\,\frac{R_{\mu\nu}}{|x-x_*'|^2}
\right],
\qquad
R_{\mu\nu}=\mathrm{diag}(1,1,1,-1),
\end{equation}
where
\begin{equation}
x_*'=(x'^1,x'^2,x'^3,-z'),
\qquad
\eta=
\begin{cases}
+1, & \text{Neumann},\\
-1, & \text{Dirichlet}.
\end{cases}
\end{equation}
We now consider the oriented Wilson line
\begin{equation}
W_q[\Sigma]=\exp\!\left(iq\int_{\Sigma} A\right),
\end{equation}
with cusp at
\begin{equation}
P=(0,0,0,y),\qquad y>0,
\end{equation}
and the two rays parametrized by
\begin{equation}
\begin{split}
X_1(s)&=(-s\cos\theta_1,0,0,y+s\sin\theta_1),\qquad s\in[0,L],\\
X_2(t)&=( t\cos\theta_2,0,0,y+t\sin\theta_2),\qquad t\in[0,L].
\end{split}
\end{equation}
The physical cusp contour is traversed along $X_1$ toward the cusp and along $X_2$ away from it.
Equivalently, using the outward parametrizations above, the oriented tangent vectors are
\begin{equation}
u_1=-\dot X_1=(\cos\theta_1,0,0,-\sin\theta_1),
\qquad
u_2=\dot X_2=(\cos\theta_2,0,0,\sin\theta_2),
\end{equation}
so that $u_1\cdot u_2=\cos(\theta_1+\theta_2).$ The free energy $F(\theta_1,\theta_2;y)\equiv -\log \langle W_q[\Sigma]\rangle$ is
\begin{equation}
F
=
\frac{q^2}{2}\int_{\Sigma} dx^\mu \int_{\Sigma} dx'^\nu\,
\langle A_\mu(x)A_\nu(x')\rangle .
\end{equation}
Subtracting the factorizing self-energy contributions of the individual rays isolates the cross-arm interaction. We furthermore adopt the standard cusp normalization in which the angular coefficient vanishes in the corresponding smooth limit, amounting to an additional angle-independent subtraction. The direct term reproduces the vacuum cusp result,
\begin{equation}
F_{\rm dir,int}
=
\frac{e^2q^2}{4\pi^2}
\left[
(\theta_1+\theta_2)\cot(\theta_1+\theta_2)-1
\right]\log\frac{L}{a},
\label{eq:maxwell_direct_result}
\end{equation}
The image contribution is
\begin{equation}
F_{\rm img,int}
=
\eta\,\frac{e^2q^2}{4\pi^2}
\left[
(\theta_1-\theta_2)\cot(\theta_1-\theta_2)-1
\right]\log\frac{L}{y},
\end{equation}
Thus the full quadratic interaction free energy is
\begin{equation}
F_{\eta,{\rm int}}
=
\frac{e^2q^2}{4\pi^2}
\left[
(\theta_1+\theta_2)\cot(\theta_1+\theta_2)-1
\right]\log\frac{L}{a}
+
\eta\,\frac{e^2q^2}{4\pi^2}
\left[
(\theta_1-\theta_2)\cot(\theta_1-\theta_2)-1
\right]\log\frac{L}{y}.
\end{equation}
To check the concavity condition on the reflection-symmetric locus $\theta_1=\theta_2=\theta$, note that
\begin{equation}
\partial_{\theta_1}\partial_{\theta_2}F_{\eta,{\rm int}}
\Big|_{\theta_1=\theta_2=\theta}
=
-\frac{e^2q^2}{2\pi^2}
\left[
m(2\theta)\log\frac{L}{a}
-
\eta\,\frac13 \log\frac{L}{y}
\right], \qquad m(x)\equiv \frac{1-x\cot x}{\sin^2 x}\,.
\end{equation}
Since $m(x)$ is increasing on $(0,\pi)$, we have $m(2\theta)\ge m(0)=1/3$. Therefore, for both the Dirichlet and Neumann cases, we have
\begin{equation}
\partial_{\theta_1}\partial_{\theta_2}F_{\eta,{\rm int}}
\Big|_{\theta_1=\theta_2}
\le 0 .
\end{equation}

In the above discussion, we have not turned on a $\vartheta$-angle. Turning on a non-zero $\vartheta$ angle gives the following action,
\begin{equation}
S
=
\int_{z\ge 0} d^4x\,
\left[
\frac{1}{4e^2}F_{\mu\nu}F^{\mu\nu}
+
\frac{i\vartheta}{32\pi^2}\epsilon_{\mu\nu\rho\sigma}F^{\mu\nu}F^{\rho\sigma}
\right],
\end{equation}
Varying the action gives
\begin{equation}
\delta S\big|_{\partial}
=
-\int_{z=0} d^3x\,
\left[
\frac{1}{e^2}F_{za}
+\frac{i\vartheta}{4\pi^2}\widetilde F_{za}
\right]\delta A_a .
\end{equation}
So, we have a new duality-rotated Neumann boundary condition,
\begin{equation}
    F_{za}+i\gamma \Tilde{F}_{za}\bigg|_{z=0}=0, \qquad \gamma=\frac{e^2 \vartheta}{4\pi^2} \in \mathbb{R}.
\end{equation}
The gauge-field propagator with the above boundary conditions takes the form\footnote{The expression for the propagator in (\ref{eq:mawellproptheta}) is consistent with the propagator for the field strength with the same boundary conditions derived in \cite{DiPietro:2019hqe}.},
\begin{equation} \label{eq:mawellproptheta}
\langle A_\mu(x)A_\nu(x')\rangle_{\gamma}
=
\frac{e^2}{4\pi^2}
\left[
\frac{\delta_{\mu\nu}}{|x-x'|^2}
+
\eta_\gamma\,\frac{R_{\mu\nu}}{|x-x_*'|^2}+4\pi^2 i \rho_\gamma \delta^a_\mu \delta^b_\nu\epsilon_{abc}\partial_{c}J(r)
\right]
\end{equation}
where $\eta_\gamma=\frac{1-\gamma^2}{1+\gamma^2}$ and $\rho_\gamma=-\frac{2\gamma}{1+\gamma^2}$. The first term is the direct propagator, the second term is the parity-even (symmetric under $\mu \xleftrightarrow{} \nu$) image propagator, and the third term is the parity-odd (antisymmetric under $\mu \xleftrightarrow{} \nu$) image propagator. Notice that this term only has a non-zero value for tangential-tangential components of the gauge fields. The indices $a,b,c$ are tangential to the boundary. The function $J(r)$ governing the parity-odd image propagator is given by
\begin{equation}
    J(r)=\frac{1}{4\pi^2 r}\tan^{-1}\left(\frac{r}{d}\right)
\end{equation}
where $d=z+z'$ is the perpendicular distance between the point $x$ and the image of $x'$. $r$ is the tangential distance. Importantly, note that the parity-odd piece does not contribute to the cusp free energy at quadratic order since its contraction with the tangent vectors along the defect rays gives zero,
\begin{equation}
    \Dot{X}^\mu_1 \Dot{X}_2^{\nu}\delta^a_\mu \delta^b_\nu \epsilon_{abc} \partial_cJ(r)=0.
\end{equation}
Intuitively, the projections of the tangent vectors to the defect rays $\Dot{X}_1, \Dot{X}_2$ onto the boundary plane are collinear, and hence their scalar triple product with another boundary-parallel vector $\partial_c J(r)$ vanishes. To calculate the free energy of the cusped defect with these boundary conditions, we just have to replace $\eta \to \eta_\gamma$ in the previous analysis. Since $|\eta_\gamma|\leq 1$, the free energy for non-zero $\vartheta$-angle interpolates between the Dirichlet and Neumann free energies and is given by
\begin{equation} 
F_{\gamma,{\rm int}}
=
\frac{e^2q^2}{4\pi^2}
\left[
(\theta_1+\theta_2)\cot(\theta_1+\theta_2)-1
\right]\log\frac{L}{a}
+
\eta_\gamma\,\frac{e^2q^2}{4\pi^2}
\left[
(\theta_1-\theta_2)\cot(\theta_1-\theta_2)-1
\right]\log\frac{L}{y}.
\end{equation}
Hence, it automatically satisfies mixed concavity simply because we have already checked that for the Dirichlet and Neumann cases,
\begin{equation} 
\partial_{\theta_1}\partial_{\theta_2}F_{\gamma,{\rm int}}
\Big|_{\theta_1=\theta_2}
\le 0 .
\end{equation}

The result for the non-planar cusp readily generalizes and has the similar structure as the free scalar example but with a different coefficient function\footnote{Note that the parity odd term in (\ref{eq:mawellproptheta}) again does not contribute because the projections of the tangent vectors to the defect rays are coplanar with the gradient vector $\partial_c J(r)$ so their scalar triple product vanishes.},
\begin{equation} \label{eq:maxwell_interaction_result}
     F_{\gamma,\rm int}
=
-\frac{e^2 q^2}{4\pi^2}\left[h(\chi_d) \log \frac{y}{a}+(h(\chi_d)+\eta_\gamma h(\chi_i))\log \frac{L}{y}\right], \qquad  h(x)\equiv \frac{\pi-x}{\tan x}+1\,.
\end{equation}
The definitions of the direct and image angles $\chi_{d,i}$ are in (\ref{eq:directimageangles}).
So, the logarithmic coefficient that characterizes the non-factorizing interaction between the cusped Wilson line and the boundary is given by
\begin{equation} \label{eq:MaxwellAIR}
    A_{\rm IR, int}=-\frac{e^2q^2}{4\pi^2}\left(h(\chi_d)+\eta_\gamma h(\chi_i)\right)\,.
\end{equation}
The UV coefficient is again given by the vacuum cusp anomalous dimension $\Gamma(\chi_d)$ as expected.
We checked numerically that the Hessians of the UV and IR logarithmic coefficients for both Dirichlet and Neumann boundary conditions ($\eta=\pm 1$) is negative semi-definite. The Hessian for the duality-rotated boundary conditions for a non-zero $\vartheta$-angle interpolates between that for the Dirichlet and Neumann boundary conditions, hence automatically satisfies negative semi-definiteness. This is evidence that the interaction free energy (\ref{eq:maxwell_interaction_result}) satisfies all the positivity constraints. In addition, just like in the scalar example, numerical tests suggest complete monotonicity in the azimuthal angle $\psi$ for the negatives of both the logarithmic coefficients as illustrated in table \ref{tab:maxwell-complete-monotonicity} for some chosen values of the parameters. In addition to verifying the above inequalities along the reflection-symmetric locus, we also numerically verified that the inequality (\ref{eq:CSAIR}) is satisfied for various angles away from the reflection-symmetric locus.

\begin{table}
\centering
\scriptsize
\setlength{\tabcolsep}{4pt}

\textbf{(a) $\Gamma(\chi)$: derivatives in the opening angle $\chi$}

\vspace{2mm}

\resizebox{\textwidth}{!}{%
\begin{tabular}{c c c c c c c c c}
\hline
$(\theta,\psi)$ & $\chi$
& $C_1$ & $C_2$ & $C_3$ & $C_4$ & $C_5$ & $C_6$ & $C_7$ \\
\hline
$(0.20,\,0.50)$ & $0.490$
& $13.9$ & $54.0$ & $328$ & $2.67{\times}10^{3}$
& $2.73{\times}10^{4}$ & $3.34{\times}10^{5}$ & $4.78{\times}10^{6}$ \\

$(0.25,\,2.70)$ & $2.48$
& $0.470$ & $0.798$ & $0.439$ & $0.949$
& $1.52$ & $4.18$ & $10.8$ \\

$(0.60,\,1.10)$ & $0.892$
& $4.52$ & $9.29$ & $29.9$ & $134$
& $748$ & $5.03{\times}10^{3}$ & $3.95{\times}10^{4}$ \\

$(0.90,\,2.20)$ & $1.17$
& $2.73$ & $4.29$ & $10.0$ & $33.9$
& $144$ & $735$ & $4.38{\times}10^{3}$ \\

$(1.25,\,1.50)$ & $0.433$
& $17.5$ & $77.8$ & $535$ & $4.94{\times}10^{3}$
& $5.70{\times}10^{4}$ & $7.89{\times}10^{5}$ & $1.28{\times}10^{7}$ \\
\hline
\end{tabular}%
}

\vspace{4mm}

\textbf{(b) $\Gamma(\chi)$: derivatives in the azimuthal angle $\psi$}

\vspace{2mm}

\resizebox{\textwidth}{!}{%
\begin{tabular}{c c c c c c c c}
\hline
$(\theta,\psi)$
& $C_1$ & $C_2$ & $C_3$ & $C_4$ & $C_5$ & $C_6$ & $C_7$ \\
\hline
$(0.20,\,0.50)$
& $13.6$ & $51.8$ & $308$ & $2.46{\times}10^{3}$
& $2.46{\times}10^{4}$ & $2.95{\times}10^{5}$ & $4.14{\times}10^{6}$ \\

$(0.25,\,2.70)$
& $0.306$ & $0.731$ & $0.265$ & $0.715$
& $0.831$ & $2.57$ & $5.20$ \\

$(0.60,\,1.10)$
& $3.52$ & $6.07$ & $15.6$ & $56.8$
& $258$ & $1.41{\times}10^{3}$ & $8.95{\times}10^{3}$ \\

$(0.90,\,2.20)$
& $0.925$ & $1.30$ & $1.20$ & $2.47$
& $5.21$ & $14.8$ & $46.1$ \\

$(1.25,\,1.50)$
& $4.14$ & $6.16$ & $11.7$ & $31.6$
& $105$ & $420$ & $1.96{\times}10^{3}$ \\
\hline
\end{tabular}%
}

\vspace{4mm}

\textbf{(c) $A_{\rm IR}$: Dirichlet boundary condition}

\vspace{2mm}

\resizebox{\textwidth}{!}{%
\begin{tabular}{c c c c c c c c}
\hline
$(\theta,\psi)$
& $C_1$ & $C_2$ & $C_3$ & $C_4$ & $C_5$ & $C_6$ & $C_7$ \\
\hline
$(0.20,\,0.50)$
& $7.05$ & $42.1$ & $309$ & $2.67{\times}10^{3}$
& $2.69{\times}10^{4}$ & $3.13{\times}10^{5}$ & $4.22{\times}10^{6}$ \\

$(0.25,\,2.70)$
& $0.0236$ & $0.0616$ & $0.0595$ & $0.177$
& $0.319$ & $1.09$ & $2.90$ \\

$(0.60,\,1.10)$
& $2.58$ & $5.83$ & $16.7$ & $59.7$
& $263$ & $1.40{\times}10^{3}$ & $8.87{\times}10^{3}$ \\

$(0.90,\,2.20)$
& $0.695$ & $1.08$ & $1.30$ & $2.66$
& $5.45$ & $15.0$ & $45.7$ \\

$(1.25,\,1.50)$
& $4.07$ & $6.16$ & $11.7$ & $31.6$
& $105$ & $420$ & $1.96{\times}10^{3}$ \\
\hline
\end{tabular}%
}

\vspace{4mm}

\textbf{(d) $A_{\rm IR}$: Neumann boundary condition}

\vspace{2mm}

\resizebox{\textwidth}{!}{%
\begin{tabular}{c c c c c c c c}
\hline
$(\theta,\psi)$
& $C_1$ & $C_2$ & $C_3$ & $C_4$ & $C_5$ & $C_6$ & $C_7$ \\
\hline
$(0.20,\,0.50)$
& $20.1$ & $61.5$ & $307$ & $2.25{\times}10^{3}$
& $2.23{\times}10^{4}$ & $2.77{\times}10^{5}$ & $4.05{\times}10^{6}$ \\

$(0.25,\,2.70)$
& $0.589$ & $1.40$ & $0.471$ & $1.25$
& $1.34$ & $4.04$ & $7.49$ \\

$(0.60,\,1.10)$
& $4.47$ & $6.32$ & $14.6$ & $53.9$
& $253$ & $1.41{\times}10^{3}$ & $9.03{\times}10^{3}$ \\

$(0.90,\,2.20)$
& $1.16$ & $1.51$ & $1.09$ & $2.28$
& $4.98$ & $14.6$ & $46.5$ \\

$(1.25,\,1.50)$
& $4.21$ & $6.16$ & $11.6$ & $31.5$
& $105$ & $420$ & $1.96{\times}10^{3}$ \\
\hline
\end{tabular}%
}

\caption{
The table illustrates the alternating-sign derivative
hierarchy for the cusp logarithms in the free Maxwell example. All angles are given in radians,
and the common positive overall factor $e^2q^2/(4\pi^2)$ is suppressed.
In (a),
$C_n\equiv(-1)^{n+1}\partial_\chi^n\Gamma(\chi)$,
with the opening angle determined by
$\cos\chi=\cos^2\theta\cos\psi+\sin^2\theta$.
In (b),
$C_n\equiv(-1)^{n+1}\partial_\psi^n
\Gamma(\chi(\theta,\psi))$
at fixed $\theta$.
In (c) and (d),
$C_n\equiv(-1)^{n+1}\partial_\psi^n A_{\rm IR}$
for Dirichlet and Neumann boundary conditions, respectively.
Notice that all the displayed signed derivatives are positive.
}
\label{tab:maxwell-complete-monotonicity}
\end{table}

Now, we verify the Lorentzian bound (\ref{eq:Lorconstraints}). We will closely follow the analysis done in the scalar example so we will skip most of the details. The Lorentzian continuations of the direct and image angles are as given in (\ref{eq:Lordirimgangles}). The analytic continuation of the function $h(x)$ to the required domain is $h(x)\to 1+zf(z)$ with the integral representation of $f(z)$ for $z<1$ given in (\ref{eq:scalarfz}). So, the Lorentzian continuation of (\ref{eq:MaxwellAIR}) is given by
\begin{equation}
     A^L_{\rm IR, int}(\theta,\rho) = -\frac{e^2q^2}{4\pi^2}(z_df(z_d)+1+\eta_\gamma(1+ z_if(z_i)))\,.
\end{equation}
For $-1\leq \eta_\gamma \leq 1$, we can check that the above function is monotonically increasing in $\rho$, 
\begin{equation}
    \frac{\partial  A^L_{\rm IR, int}(\theta,\rho) }{\partial \rho} \geq 0, \qquad \rho \geq 0\,.
\end{equation}
So, the Lorentzian bound is satisfied,
\begin{equation}
    A^L_{\rm IR, int}(\theta,\rho) - A^L_{\rm IR, int}(\theta,\rho=0) \geq 0\,.
\end{equation}
We have therefore verified all the positivity constraints for the Maxwell case as well.

\subsection{Planar \texorpdfstring{$\mathcal N=4$}{N=4} SYM at weak coupling} \label{sec:SYMweak}

In the previous two examples, we have computed the logarithmic coefficients and verified the positivity constraints in free theories.
We will now verify the constraints in an interacting theory perturbatively, for a cusped Maldacena-Wilson line in planar \(\mathcal N=4\) SYM at weak 't Hooft coupling on the half-space $\mathbb H^4$. We write the six real scalars as
\begin{equation}
\Phi_I=(X_1,X_2,X_3,Y_1,Y_2,Y_3).
\end{equation}
The Euclidean bosonic action is\footnote{We use Hermitian generators normalized by $\text{tr}(T^aT^b)=\frac{1}{2}\delta^{ab}$.}
\begin{equation}
S_{\rm bos}
=
\frac{1}{g_{\rm YM}^2}
\int_{z\ge0} d^4x\,
\operatorname{tr}\left[
\frac12 F_{\mu\nu}F^{\mu\nu}
+
D_\mu\Phi_I D^\mu\Phi_I
-\frac12[\Phi_I,\Phi_J]^2
\right].
\end{equation}
We impose the NS5-like boundary conditions \cite{GaiottoWitten2009Boundary},
\begin{equation}
F_{za}\big|_{z=0}=0,\qquad D_z X_i\big|_{z=0}=0,
\qquad Y_i\big|_{z=0}=0,
\qquad a=1,2,3.
\end{equation}
Thus the tangential gauge field \(A_a\) and the triplet \(X_i\) obey Neumann boundary
conditions while the triplet \(Y_i\) obeys Dirichlet boundary conditions. In addition, in the Feynman gauge description used below, we supplement these boundary conditions with the compatible Dirichlet boundary condition $A_z\big|_{z=0}=0$.

We consider the standard $1/2$-BPS Maldacena-Wilson line \cite{Maldacena1998WilsonLoops} in the fundamental representation, 
\begin{equation} \label{eq:MaldacenaWilson}
W[\Sigma]
=
\frac1N\operatorname{tr}\,P\exp\int_\Sigma d\tau\,
\left(
i\dot x^\mu A_\mu
+
|\dot x|\, n^iX_i
\right),
\qquad
n^i n^i=1,
\end{equation}
where \(n^i\) is chosen inside the Neumann scalar triplet \(X_i\).
The half-space propagators relevant for the Wilson line are
\begin{align}
\langle A_\mu^a(x)A_\nu^b(x')\rangle
&=
\frac{g_{\rm YM}^2\delta^{ab}}{4\pi^2}
\left[
\frac{\delta_{\mu\nu}}{|x-x'|^2}
+
\frac{R_{\mu\nu}}{|x-x^{\prime *}|^2}
\right],
\\
\langle X_i^a(x)X_j^b(x')\rangle
&=
\frac{g_{\rm YM}^2\delta^{ab}\delta_{ij}}{4\pi^2}
\left[
\frac{1}{|x-x'|^2}
+
\frac{1}{|x-x^{\prime *}|^2}
\right].
\end{align}
where $x^{\prime *}=(x^{\prime 1},x^{\prime 2},x^{\prime 3},-z')$ and $R_{\mu\nu}=\operatorname{diag}(1,1,1,-1).$
We use the same parametrization for the location of the cusp and the orientation of the defect rays as used in the previous two examples.

At leading order in the planar limit, \(\lambda=g_{\rm YM}^2N\), 
adding the direct and image contributions to the free energy \(F=-\log\langle W\rangle\) for a planar cusp, we get
\begin{align}
F^{(1)}
=
-&\frac{\lambda}{8\pi^2}
(\theta_1+\theta_2)\tan\frac{\theta_1+\theta_2}{2}
\log\frac{L}{a}
\nonumber\\
-&\frac{\lambda}{16\pi^2}
\left[
(\pi-2\theta_1)\tan\theta_1
+
(\pi-2\theta_2)\tan\theta_2
+
2(\theta_1-\theta_2)\tan\frac{\theta_1-\theta_2}{2}
\right]
\log\frac{L}{y}.
\end{align}
As in the scalar and Maxwell examples, we can subtract the free energies of the two isolated
rays in the same half-space. This gives the interaction part of the free energy,
\begin{equation}
F^{(1)}_{\rm int}
=
-\frac{\lambda}{8\pi^2}
\left[
(\theta_1+\theta_2)\tan\left(\frac{\theta_1+\theta_2}{2}\right)\log\frac{L}{a}
+
(\theta_1-\theta_2)\tan\left(\frac{\theta_1-\theta_2}{2}\right)\log\frac{L}{y}
\right].
\end{equation}
On the reflection-symmetric locus \(\theta_1=\theta_2\), the mixed second derivative of the above expression is consistent with concavity,
\begin{equation}
\left.
\partial_{\theta_1}\partial_{\theta_2}F^{(1)}_{\rm int}
\right|_{\theta_1=\theta_2}
\le 0.
\end{equation}
Thus the Maldacena-Wilson cusp in \(\mathcal N=4\) SYM with NS5-like boundary
conditions satisfies the mixed concavity condition at weak coupling.

\begin{table}
\centering
\scriptsize
\setlength{\tabcolsep}{4pt}

\textbf{(a) $\Gamma(\chi)$: derivatives in the opening angle $\chi$}

\vspace{2mm}

\resizebox{\textwidth}{!}{%
\begin{tabular}{c c c c c c c c c}
\hline
$(\theta,\psi)$ & $\chi$
& $C_1$ & $C_2$ & $C_3$ & $C_4$ & $C_5$ & $C_6$ & $C_7$ \\
\hline
$(0.20,\,0.50)$ & $0.490$
& $26.6$ & $107$ & $655$ & $5.35{\times}10^{3}$
& $5.46{\times}10^{4}$ & $6.69{\times}10^{5}$ & $9.56{\times}10^{6}$ \\

$(0.25,\,2.70)$ & $2.48$
& $0.716$ & $1.25$ & $0.831$ & $1.82$
& $3.01$ & $8.29$ & $21.6$ \\

$(0.60,\,1.10)$ & $0.892$
& $8.14$ & $18.0$ & $59.5$ & $267$
& $1.50{\times}10^{3}$ & $1.01{\times}10^{4}$ & $7.90{\times}10^{4}$ \\

$(0.90,\,2.20)$ & $1.17$
& $4.71$ & $8.08$ & $19.8$ & $67.6$
& $288$ & $1.47{\times}10^{3}$ & $8.76{\times}10^{3}$ \\

$(1.25,\,1.50)$ & $0.433$
& $33.9$ & $155$ & $1.07{\times}10^{3}$ & $9.88{\times}10^{3}$
& $1.14{\times}10^{5}$ & $1.58{\times}10^{6}$ & $2.55{\times}10^{7}$ \\
\hline
\end{tabular}%
}

\vspace{4mm}

\textbf{(b) $\Gamma(\chi)$: derivatives in the azimuthal angle $\psi$}

\vspace{2mm}

\resizebox{\textwidth}{!}{%
\begin{tabular}{c c c c c c c c}
\hline
$(\theta,\psi)$
& $C_1$ & $C_2$ & $C_3$ & $C_4$ & $C_5$ & $C_6$ & $C_7$ \\
\hline
$(0.20,\,0.50)$
& $26.0$ & $103$ & $615$ & $4.92{\times}10^{3}$
& $4.92{\times}10^{4}$ & $5.91{\times}10^{5}$ & $8.27{\times}10^{6}$ \\

$(0.25,\,2.70)$
& $0.466$ & $1.13$ & $0.502$ & $1.36$
& $1.64$ & $5.08$ & $10.4$ \\

$(0.60,\,1.10)$
& $6.34$ & $11.7$ & $31.1$ & $113$
& $516$ & $2.81{\times}10^{3}$ & $1.79{\times}10^{4}$ \\

$(0.90,\,2.20)$
& $1.59$ & $2.31$ & $2.36$ & $4.91$
& $10.4$ & $29.5$ & $92.2$ \\

$(1.25,\,1.50)$
& $8.00$ & $12.2$ & $23.3$ & $63.1$
& $210$ & $840$ & $3.92{\times}10^{3}$ \\
\hline
\end{tabular}%
}

\vspace{4mm}

\textbf{(c) $A_{\rm IR}$: $\eta_\gamma=-1$}

\vspace{2mm}

\resizebox{\textwidth}{!}{%
\begin{tabular}{c c c c c c c c}
\hline
$(\theta,\psi)$
& $C_1$ & $C_2$ & $C_3$ & $C_4$ & $C_5$ & $C_6$ & $C_7$ \\
\hline
$(0.20,\,0.50)$
& $25.2$ & $103$ & $619$ & $4.94{\times}10^{3}$
& $4.93{\times}10^{4}$ & $5.91{\times}10^{5}$ & $8.27{\times}10^{6}$ \\

$(0.25,\,2.70)$
& $0.328$ & $0.810$ & $0.485$ & $1.33$
& $1.66$ & $5.15$ & $10.6$ \\

$(0.60,\,1.10)$
& $6.00$ & $11.7$ & $31.6$ & $114$
& $516$ & $2.81{\times}10^{3}$ & $1.79{\times}10^{4}$ \\

$(0.90,\,2.20)$
& $1.48$ & $2.22$ & $2.44$ & $5.01$
& $10.5$ & $29.5$ & $91.9$ \\

$(1.25,\,1.50)$
& $7.96$ & $12.2$ & $23.4$ & $63.1$
& $210$ & $840$ & $3.92{\times}10^{3}$ \\
\hline
\end{tabular}%
}

\vspace{4mm}

\textbf{(d) $A_{\rm IR}$: $\eta_\gamma=+1$}

\vspace{2mm}

\resizebox{\textwidth}{!}{%
\begin{tabular}{c c c c c c c c}
\hline
$(\theta,\psi)$
& $C_1$ & $C_2$ & $C_3$ & $C_4$ & $C_5$ & $C_6$ & $C_7$ \\
\hline
$(0.20,\,0.50)$
& $38.2$ & $122$ & $617$ & $4.52{\times}10^{3}$
& $4.47{\times}10^{4}$ & $5.55{\times}10^{5}$ & $8.10{\times}10^{6}$ \\

$(0.25,\,2.70)$
& $0.893$ & $2.15$ & $0.896$ & $2.40$
& $2.69$ & $8.10$ & $15.2$ \\

$(0.60,\,1.10)$
& $7.88$ & $12.2$ & $29.5$ & $108$
& $506$ & $2.82{\times}10^{3}$ & $1.80{\times}10^{4}$ \\

$(0.90,\,2.20)$
& $1.94$ & $2.65$ & $2.22$ & $4.63$
& $10.0$ & $29.1$ & $92.7$ \\

$(1.25,\,1.50)$
& $8.10$ & $12.2$ & $23.2$ & $63.1$
& $210$ & $840$ & $3.92{\times}10^{3}$ \\
\hline
\end{tabular}%
}

\caption{
The table illustrates the alternating-sign derivative
hierarchy for the cusp logarithms in planar $\mathcal N=4$ SYM at weak coupling.
All angles are given in radians, and the common positive overall factor
$\lambda/(8\pi^2)$ is suppressed.
In (a),
$C_n\equiv(-1)^{n+1}\partial_\chi^n\Gamma(\chi)$,
with the opening angle determined by
$\cos\chi=\cos^2\theta\cos\psi+\sin^2\theta$.
In (b),
$C_n\equiv(-1)^{n+1}\partial_\psi^n
\Gamma(\chi(\theta,\psi))$
at fixed $\theta$.
In (c) and (d),
$C_n\equiv(-1)^{n+1}\partial_\psi^n A_{\rm IR}$
for the endpoint values $\eta_\gamma=-1$ and
$\eta_\gamma=+1$, respectively.
Notice that all the displayed signed derivatives are positive.
}
\label{tab:sym-complete-monotonicity}
\end{table}

We now turn on a constant Yang–Mills theta angle. For the supersymmetric NS5-like boundary condition, the scalar Neumann/Dirichlet assignment remains unchanged while the gauge-field boundary condition is duality rotated \cite{Gaiotto:2008sd}. The action for the gauge field is modified as
\begin{equation}
S_{\rm gauge}
=
\int_{z\ge 0} d^4x\,
\text {tr}\left[
\frac{1}{2 g_{\rm YM}^2}F_{\mu\nu}F^{\mu\nu}
+
\frac{i\vartheta}{32\pi^2}\epsilon_{\mu\nu\rho\sigma}F^{\mu\nu}F^{\rho\sigma}
\right],
\end{equation}
while the action for the scalars is unchanged. Varying the action gives the duality-rotated boundary condition
\begin{equation}
    F_{za}+i\gamma \Tilde{F}_{za}\bigg|_{z=0}=0, \qquad \gamma=\frac{g_{\rm YM}^2 \vartheta}{8\pi^2} \in \mathbb{R}.
\end{equation}
The gauge-field propagator with the above boundary conditions takes the form,
\begin{equation} \label{eq:mawellpropagatortheta}
\langle A^a_\mu(x)A^b_\nu(x')\rangle_{\gamma}
=
\frac{g_{\rm YM}^2 \delta^{ab}}{4\pi^2}
\left[
\frac{\delta_{\mu\nu}}{|x-x'|^2}
+
\eta_\gamma\,\frac{R_{\mu\nu}}{|x-x_*'|^2}+4\pi^2 i \rho_\gamma \delta^i_\mu \delta^j_\nu\epsilon_{ijk}\partial_{k}J(r)
\right]\,.
\end{equation}
with the same notation as in the Maxwell case with $\eta_\gamma=\frac{1-\gamma^2}{1+\gamma^2}$ and $\rho_\gamma=-\frac{2\gamma}{1+\gamma^2}$. The scalar propagators are unchanged. So, the interaction part of the cusp free energy takes the form\footnote{As in the Maxwell example, the parity-odd term in the gauge-field propagator does not contribute to the free energy due to coplanarity. The same is true for the non-planar cusp as well.} 
\begin{equation}
F^{(1)}_{\gamma,\rm int}
=
-\frac{\lambda}{8\pi^2}
\left[
(\theta_1+\theta_2)\tan\left(\frac{\theta_1+\theta_2}{2}\right)\log\frac{L}{a}
+
(1-\eta_\gamma \cos (\theta_1-\theta_2))\frac{\theta_1-\theta_2}{\sin(\theta_1-\theta_2)}\log\frac{L}{y}
\right],
\end{equation}
On the reflection-symmetric locus \(\theta_1=\theta_2\), it is easy to see that the mixed second derivative is consistent with concavity,
\begin{equation}
    \left.
\partial_{\theta_1}\partial_{\theta_2}F^{(1)}_{\gamma,\rm int}
\right|_{\theta_1=\theta_2}\leq 0.
\end{equation}
For the general non-planar cusp, we can write down a similar expression for the free energy in terms of the direct and image angles that we defined earlier and separating into the UV and IR contributions,
\begin{equation}
    F^{(1)}_{\gamma,\rm int}
= -\frac{\lambda}{8\pi^2}\left[G(\chi_d)\log\frac{y}{a}+(G(\chi_d)+H_{\eta_\gamma}(\chi_i))\log\frac{L}{y}\right]
\end{equation}
where the two coefficient functions are given by
\begin{equation}
    G(x)=\frac{\pi-x}{\tan \frac{x}{2}}, \qquad H_\eta(x)=(1+\eta \cos x)\frac{\pi-x}{\sin x}
\end{equation}
The angles $\chi_d$ and $\chi_i$ are defined as in (\ref{eq:directimageangles}).
So, the logarithmic coefficient that characterizes the non-factorizing interaction of the Maldacena-Wilson line with the boundary is given by
\begin{equation} \label{eq:AIRSYM}
    A_{\rm IR, int}=-\frac{\lambda}{8\pi^2}\left(G(\chi_d)+H_{\eta_\gamma}(\chi_i)\right)\,.
\end{equation}
The UV coefficient is the vacuum cusp anomalous dimension $\Gamma(\chi_d)$ as expected.
We have verified numerically that both the logarithmic coefficients obey all the positivity constraints for $|\eta_\gamma|\leq 1$. In addition, just like with the scalar and Maxwell examples, numerical tests suggest that the negatives of these coefficients are completely monotonic functions of the azimuthal angle $\psi$ as illustrated in table \ref{tab:sym-complete-monotonicity} for some chosen values of the parameters. In addition to verifying the above inequalities along the reflection-symmetric locus, we also numerically verified that the inequality (\ref{eq:CSAIR}) is satisfied for various angles away from the reflection-symmetric locus. Finally, to verify the Lorentzian constraint (\ref{eq:Lorconstraints}), note that the Lorentzian continuation of (\ref{eq:AIRSYM}) is given by
\begin{equation}
    A^L_{\rm IR, int}(\theta, \rho)= -\frac{\lambda}{8\pi^2}\left(f(z_d)(1+z_d)+f(z_i)(1+\eta_\gamma z_i)\right)\,.
\end{equation}
The parameters $z_d,z_i$ are as defined in (\ref{eq:Lordirimgangles}) and the function $f(z)$ is given by the integral representation (\ref{eq:scalarfz}). It is straightforward to check that the above Lorentzian coefficient is monotonically increasing in $\rho$ for $\rho \geq 0$ hence the bound (\ref{eq:Lorconstraints}) is satisfied. We have therefore verified that all the positivity constraints are satisfied in this setup.

\section{The holographic cusp in the strongly-coupled D3-D5 defect CFT} \label{sec:holographiccusp}

In this section, we turn to the strongly coupled D3-D5 defect CFT, where the cusped Maldacena-Wilson line used in Sec \ref{sec:SYMweak} is described holographically by a classical string worldsheet. In this example there are two competing worldsheet saddles, one ending on the D5-brane and the other corresponding to the vacuum cusp, which undergo a Gross-Ooguri-type phase transition as the cusp angles are varied. This provides a non-trivial test that the positivity constraints continue to hold not only within each phase, but also across the transition between them.

We now study the cusped fundamental Maldacena-Wilson line (\ref{eq:MaldacenaWilson}) at strong coupling in the
D3-D5 defect CFT \cite{KarchRandall2001,DeWolfeFreedmanOoguri2002}. In the planar large-$N$ and large-'t Hooft coupling
limit, the expectation value of the Wilson line is determined by the
renormalized area of a classical fundamental string worldsheet \cite{Maldacena1998WilsonLoops, Rey:1998ik},
\begin{equation}
    F[C]
    \equiv
    -\log \langle W[C]\rangle
    =
    S_{\mathrm{ren}}[C]
    +O(\lambda^0),
    \qquad
    S
    =
    \frac{\sqrt{\lambda}}{2\pi}
    \int_{\Sigma}d^2\sigma\,
    \sqrt{\det g_{ab}} .
\end{equation}
Here and below the AdS radius is set to one. We use Euclidean Poincar\'e
coordinates
\begin{equation}
    ds^2_{\mathrm{AdS}_5}
    =
    \frac{
    dz^2+dx^2+dx_\perp^2+d\mathbf{x}_{\parallel}^{\,2}
    }{z^2},
    \qquad z>0,
\end{equation}
where $x$ is the coordinate along
the undeformed Wilson line and $x_\perp$ is the boundary direction normal to the $3d$ interface that sources the probe D5-brane in the bulk. The interface is at $x_\perp=0$ on the boundary, $z=0$. The remaining two directions denoted $\mathbf{x}_{\parallel}$ are parallel to the interface. The probe D5-brane has worldvolume AdS$_4 \times S^2 \subset$ AdS$_5\times S^5$, with AdS embedding
\begin{equation}
    x_\perp=\kappa z,
    \qquad
    \kappa=\frac{\pi k}{\sqrt{\lambda}}.
\end{equation}
The $k \in \mathbb{Z}_{\geq 0}$ measures the jump in the rank of the gauge group across the interface and equals to the quantized worldvolume flux through the wrapped $S^2$. Throughout this section we take $\kappa \geq 0$.

\subsection{The worldsheet action and boundary conditions}

We now derive the bulk equations for the worldsheet and its boundary conditions at the D5-brane by varying the Nambu-Goto action. Restricting
to the $\mathrm{AdS}_3$ subspace parametrized by $(z,x,x_\perp)$ gives
\begin{equation}
    S
    =
    \frac{\sqrt{\lambda}}{2\pi}
    \int d^2\sigma\,
   \frac{\sqrt{\text{det}\left(\partial_a z\partial_b z+ \partial_a x\partial_b x+\partial_a x_\perp \partial_b x_\perp\right)}}{z^2}.
\end{equation}
Here, the scalar polarization $n^I$ is chosen to select a point on the $S^2\subset S^5$ wrapped by the D5-brane. For this choice, the worldsheet saddle may be taken at fixed $S^5$ position, so only the AdS$_3$ coordinates enter the Nambu-Goto action. Varying the Nambu-Goto action and choosing conformal worldsheet coordinates, the bulk equations for the three non-trivial target-space coordinates are
\begin{equation} \label{eq:eom}
\begin{split}
    z\,\partial^2x
    -2\,\partial_a z\,\partial_a x&=0,
    \\
    z\,\partial^2x_\perp
    -2\,\partial_a z\,\partial_a x_\perp&=0,
    \\
    z\,\partial^2z
    -(\partial z)^2
    +(\partial x)^2
    +(\partial x_\perp)^2&=0.
    \end{split}
\end{equation}
The conformal-gauge condition gives the Virasoro constraints
\begin{equation} \label{eq:Virasoro}
\begin{split}
    (\partial_1z)^2+(\partial_1x)^2+(\partial_1x_\perp)^2
    &=
    (\partial_2z)^2+(\partial_2x)^2+(\partial_2x_\perp)^2,
    \\
    \partial_1z\,\partial_2z
    +\partial_1x\,\partial_2x
    +\partial_1x_\perp\,\partial_2x_\perp
    &=0.
    \end{split}
\end{equation}
Equivalently, the same equations follow from the Polyakov action in conformal gauge.

At the regulated AdS boundary, the string endpoint is fixed on the Wilson
line contour:
\begin{equation}
    z=\epsilon,
    \qquad
    (x,x_\perp)=C_{\mathrm{cusp}}.
\end{equation}
At the other end, the string endpoint is constrained to lie on the D5
brane but is free to move along it. The boundary term in the variation is
\begin{equation}
    \delta S_{\partial\Sigma}
    =
    \frac{\sqrt{\lambda}}{2\pi}
    \int_{\partial\Sigma_{\mathrm{D5}}}ds\,
    \frac{1}{z^2}
    \left(
        \partial_nz\,\delta z
        +\partial_nx\,\delta x
        +\partial_nx_\perp\,\delta x_\perp
    \right),
\end{equation}
where $\partial_n$ is the outward normal derivative on the worldsheet.
Allowed endpoint variations are tangent to the D5-brane and hence obey
\begin{equation}
    \delta x_\perp=\kappa\,\delta z.
\end{equation}
Requiring the boundary variation to vanish for arbitrary tangent
variations gives
\begin{equation} \label{eq:D5bdryconditions}
    x_\perp=\kappa z,
    \qquad
    \partial_nx=0,
    \qquad
    \partial_nz+\kappa\,\partial_nx_\perp=0.
\end{equation}
In particular, the string meets the D5-brane orthogonally in the $(x_\perp,z)$-plane, in agreement with the boundary conditions used in \cite{NagasakiTanidaYamaguchi2012}.
The first condition fixes the endpoint to the D5-brane, while the
remaining conditions are the Neumann conditions along the brane. In
particular, the string meets the D5-brane orthogonally in the
$(z,x_\perp)$ plane.

\subsection{Review: The holographic interface-particle potential}

We first review the solution without a cusp. The straight-line solution and the corresponding strong-coupling interface-particle potential were obtained in \cite{NagasakiTanidaYamaguchi2012}\footnote{Perturbative corrections to the same straight Wilson line and their comparison with the worldsheet calculation in the large-$k$ double-scaling limit were subsequently studied in \cite{deLeeuw:2016vgp}.}; we briefly review the calculation in conventions adapted to the cusp problem. Consider an infinite straight
line parallel to the interface and located at $x_\perp=y$. We use the
ansatz
\begin{equation}
    x=\tau,
    \qquad
    x_\perp=Y_0(\sigma),
    \qquad
    z=Z_0(\sigma),
\end{equation}
and choose $\sigma$ to be arclength in the $(Y_0,Z_0)$ plane, $(Y_0')^2+(Z_0')^2=1.$
The reduced Nambu-Goto action is
\begin{equation}
    S
    =
    \frac{\sqrt{\lambda}}{2\pi}
    \int d\tau\,d\sigma\,\frac{1}{Z_0^2}.
\end{equation}
Translation invariance in $Y_0$ gives a conserved quantity,
\begin{equation}
    Y_0'=-cZ_0^2
    \implies
    Z_0'=\sqrt{1-c^2Z_0^4}\,,
\end{equation}
where $c>0$ is a constant.
The boundary conditions are
\begin{equation}
    Z_0(0)=0,
    \qquad
    Y_0(0)=y,
\end{equation}
at the AdS boundary, and
\begin{equation}
    Y_0(\sigma_\ast)=\kappa Z_0(\sigma_\ast),
    \qquad
    Z_0'(\sigma_\ast)+\kappa Y_0'(\sigma_\ast)=0,
\end{equation}
at the D5-brane. Introducing $u=\sqrt{c}\,Z_0$, the D5 endpoint is located at $u_\ast=(1+\kappa^2)^{-1/4}.$
The relation between $c$ and the distance $y$ is
\begin{equation}
    \sqrt{c}\,y=A_\kappa,
    \qquad
    A_\kappa
    =
    \int_0^{u_\ast}
    du\,\frac{u^2}{\sqrt{1-u^4}}
    +\kappa u_\ast.
\end{equation}
After subtracting the standard Wilson-line perimeter divergence at $z=\epsilon$ \cite{Drukker:1999zq}, the
renormalized action per unit length $L$ is
\begin{equation} \label{eq:stline}
    \frac{S_{\mathrm{ren}}}{L}
    =
    -\frac{\sqrt{\lambda}}{2\pi}
    \frac{A_\kappa^2}{y}.
\end{equation}
This is the strong-coupling interface-particle potential \cite{NagasakiTanidaYamaguchi2012}. At large $\kappa$, the coefficient of the potential $A_\kappa^2$ has an expansion of the form
\begin{equation} \label{eq:stlinelargekappa}
    A_\kappa^2=\kappa+\frac{1}{6\kappa}-\frac{19}{504\kappa^3}+O(\kappa^{-5}).
\end{equation}
This expansion will be useful when we generalize the interface-particle potential to a cusped line below.

\subsection{The cusp free energy and a phase transition}

For a planar cusp, the two rays at the AdS boundary are parametrized as
\begin{equation}
\begin{split}
    (x,x_\perp)_1
    &=
    \left(-s\cos\theta_1,\,
    y+s\sin\theta_1\right),
    \qquad s\geq 0,
    \\
    (x,x_\perp)_2
    &=
    \left(s\cos\theta_2,\,
    y+s\sin\theta_2\right),
    \qquad s\geq 0.
    \end{split}
\end{equation}
The coefficients of the logarithms $\log(L/y)$ and $\log(y/a)$ are respectively controlled by the regimes $s\gg y$ and $s \ll y$. Below, we will describe the geometry of the worldsheet corresponding to these regimes and calculate the classical action. We will ignore the crossover region $s\sim y$ in the subsequent analysis.

\begin{figure}
    \centering
    \includegraphics[width=0.7\linewidth]{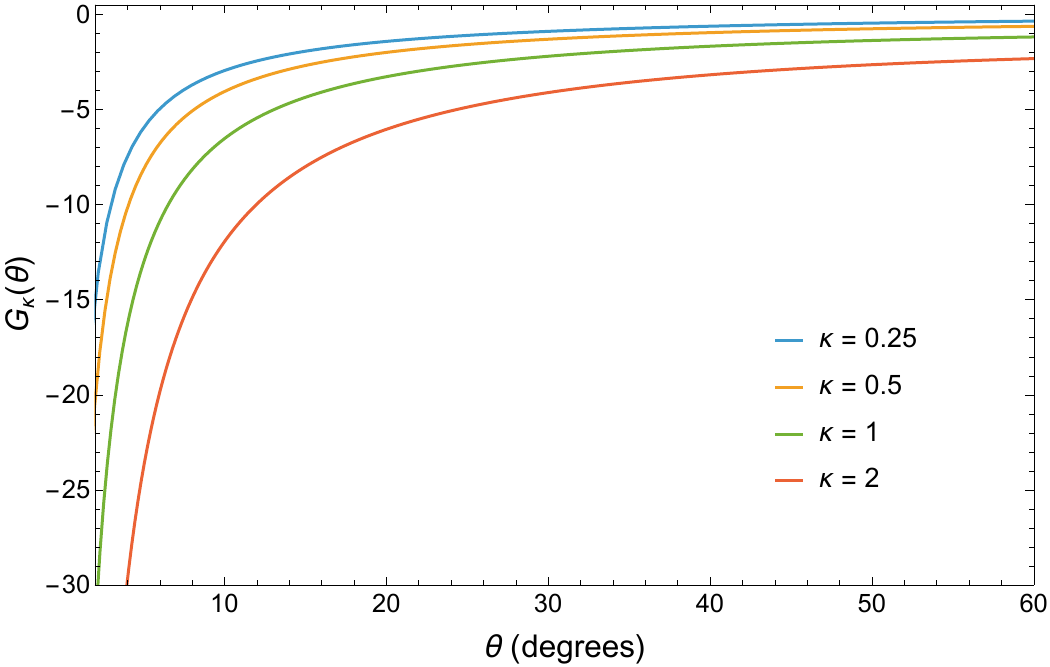}
    \caption{This is a graph of the function $G_\kappa(\theta)$ that governs the coefficient of the IR cusp logarithm in the D5-ending phase. It can be also be interpreted as the interface-particle potential for an oblique defect ray making an angle $\theta$ with the interface. We have plotted the function for different flux parameters $\kappa$. The curves approach the limiting value $-\kappa$ as $\theta \to \frac{\pi}{2}$ consistent with (\ref{eq:largeanglesGkappa}).}
    \label{fig:potentialcuspD5}
\end{figure}

In the $s\gg y$ regime used to compute the IR cusp logarithm, we can ignore the separation $y$ between the cusp and the interface and therefore compute the area of the worldsheet stretched between the AdS boundary and the D5-brane using a scale-invariant ansatz,
\begin{equation} \label{eq:embed}
    x=R\cos(\phi), \quad x_\perp= R\sin(\phi), \quad z=Rh(\phi)\,.
\end{equation}
Here, $(R,\phi)$ are the worldsheet coordinates. The boundary regime $s\gg y$ translates to $R\gg y$. In this regime, the contribution from each defect ray factorizes, so we just have to add the two contributions from the worldsheet area to get the IR cusp logarithm. The angular coordinate of the worldsheet $\phi$ extends between $(\theta,\phi_0)$ which are respectively the values at the AdS boundary and the D5-brane. $\theta$ is a shorthand for either cusp angles $\theta_1$ or $\theta_2$. The embedding function $h(\phi)$ satisfies $h(\theta)=0$ at the AdS boundary and $h(\phi_0)\equiv h_0$ at the D5-brane. The constants $\phi_0$ and $h_0$ will be determined below by solving for the geometry of the worldsheet. Using the metric induced on the worldsheet by its embedding (\ref{eq:embed}) into AdS$_3$,
\begin{equation}
    ds^2_\text{ind}=\frac{1}{h^2}\left[(1+h^2)\frac{dR^2}{R^2}+2hh'\frac{dR}{R}d\phi+(1+h'^2)d\phi^2\right]\,,
\end{equation}
the worldsheet action sourced by the defect ray is
\begin{equation} \label{eq:defectarmaction}
    S_h=\frac{\sqrt{\lambda}}{2\pi}\int \frac{dR}{R}\int_{\phi_0}^{\theta}d\phi \mathcal{L}_h, \qquad \mathcal{L}_h\equiv\frac{\sqrt{1+h^2+h'^2}}{h^2}\,.
\end{equation}
The $R$ integral gives the IR logarithm $\log(L/y)$. The D5 boundary conditions (\ref{eq:D5bdryconditions}) translate to
\begin{equation} \label{eq:bdryconditionsD5arm}
    \sin(\phi_0)=\kappa h_0, \quad h'(\phi_0)=-\frac{\kappa(1+h_0^2)}{\cos \phi_0}\,.
\end{equation}
The first condition arises from requiring that the endpoint is on the D5-brane and the second condition follows from the Neumann conditions along the brane which together ensure that the variational problem associated with (\ref{eq:defectarmaction}) is well-posed. To solve the Euler-Lagrange equations for the Lagrangian $\mathcal{L}_h$, it is useful to identify that since it has no explicit $\phi$ dependence, there is an integral of motion given by
\begin{equation} \label{eq:equationofmotion}
    C\equiv \mathcal{L}_h - \frac{\partial \mathcal{L}_h}{\partial h'}h'=\frac{1+h^2}{h^2\sqrt{1+h^2+h'^2}}\,.
\end{equation}
Using (\ref{eq:bdryconditionsD5arm}), we can express the constant $C$ as 
\begin{equation}
    C=\frac{\cos \phi_0\sqrt{1+h_0^2}}{h_0^2\sqrt{1+\kappa^2}}, \qquad \phi_0=\sin^{-1}(\kappa h_0)\,.
\end{equation}
Integrating the equation of motion (\ref{eq:equationofmotion}) from the AdS boundary to the D5-brane, we get
\begin{equation}
    \theta=\phi_0+\int_0^{h_0}dh \frac{Ch^2}{\sqrt{1+h^2}\sqrt{1+h^2-C^2 h^4}}\,.
\end{equation}
This determines the remaining independent constant $h_0$ in terms of the cusp angle $\theta$. Having determined the geometry, we now compute the action of the worldsheet,
\begin{equation}
    S=\frac{\sqrt{\lambda}}{2\pi}\log\frac{L}{y}\int_0^{h_0}dh\frac{\sqrt{1+h^2}}{h^2\sqrt{1+h^2-C^2h^4}}\,.
\end{equation}
The integral diverges linearly near the AdS boundary so holographic renormalization is necessary. We define the renormalized integral as
\begin{equation}
    G_\kappa(\theta)\equiv \lim_{\epsilon \to 0}\left[\int_\epsilon^{h_0}dh\frac{\sqrt{1+h^2}}{h^2\sqrt{1+h^2-C^2h^4}}-\frac{1}{\epsilon}\right]\,.
\end{equation}
After cancelling the divergence, the above renormalized expression can be expressed in the manifestly convergent form,
\begin{equation}
    G_{\kappa}(\theta)=-\frac{1}{h_0}+\int_0^{h_0}dh \frac{C^2 h^2}{\sqrt{Q(h)}(\sqrt{1+h^2}+\sqrt{Q(h)})}, \qquad Q(h)\equiv 1+h^2-C^2 h^4.
\end{equation}
Numerically, we observe that $G_\kappa(\theta)<0$ for various values of $\kappa$ as seen in the plot in figure \ref{fig:potentialcuspD5}. In addition, we find strong numerical evidence that $-G_\kappa(\theta)$ is a completely monotonic function in $\theta$,
\begin{equation} \label{eq:completemonotonicity}
    (-1)^{n+1}\frac{d^n G_\kappa(\theta)}{d\theta^n}>0, \qquad \theta \in (0,\frac{\pi}{2})\,,
\end{equation}
as illustrated in the table (\ref{tab:holographic-complete-monotonicity}) for some chosen values of the parameters.
Now, we compute the function $G_\kappa(\theta)$ in various limits. At large $\kappa$, it admits an expansion in powers of $\frac{1}{\kappa^2}$,
\begin{equation} \label{eq:largekappaG}
    G_\kappa(\theta)=-\frac{\kappa}{\sin \theta}-\frac{\cos^2\theta}{6\kappa \sin \theta}+\frac{\cos^2\theta(112-17 \cos^2\theta)}{2520\kappa^3\sin \theta}+O(\kappa^{-5})\,.
\end{equation}
We expect the terms in this expansion can be reproduced from a planar gauge theory computation in a double scaling limit discussed in \cite{Nagasaki:2012re,NagasakiTanidaYamaguchi2012, deLeeuw:2016vgp} where $\lambda, k \to \infty$ with $\frac{\lambda}{k^2}$ fixed. For example, the leading term just follows from the tree level computation of the expectation value of the defect ray at angle $\theta$ with the interface, in a fuzzy-funnel background \cite{Constable:1999ac}, giving the free energy $F_{\rm tree}=-\frac{k-1}{2\sin \theta}\log\frac{L}{y}$ for an oblique defect ray.

Now, we evaluate $G_\kappa(\theta)$ at small and large angles. In the $\theta \to 0$ limit, we expect it to reproduce the straight-line result (\ref{eq:stline}) and indeed, we have verified numerically that
\begin{equation}
    -\lim_{\theta \to 0}\theta G_{\kappa}(\theta)= A_\kappa^2\,.
\end{equation}
The large $\kappa$ expansions of $A_\kappa^2$ in (\ref{eq:stlinelargekappa}) and $G_\kappa(\theta)$ in (\ref{eq:largekappaG}) provide an additional analytic check. We can therefore view it as a generalization of the interface-particle potential to non-zero cusp angles. At the other extreme when $\theta \to \frac{\pi}{2}$, the function approaches a constant value, 
\begin{equation} \label{eq:largeanglesGkappa}
    G_\kappa(\theta)=-\kappa-\frac{\left(\frac{\pi}{2}-\theta\right)^2}{2\cot^{-1}(\kappa)}+O\left(\left(\frac{\pi}{2}-\theta\right)^4\right), \qquad \theta\to \frac{\pi}{2}\,.
\end{equation}
We can similarly compute the free energy contribution from the complementary $s\ll y$ regime where the role of the D5-brane can be ignored and we recover the vacuum cusp anomalous dimension computed in \cite{Drukker:1999zq} which we review below. Summing up, the free energy of the holographic cusp computed using the worldsheet ending on the D5-brane is given by
\begin{equation} \label{eq:D5freeenergy}
    F_{D5}=\Gamma(\pi-\theta_1-\theta_2)\log\frac{y}{a}+\frac{\sqrt{\lambda}}{2\pi}(G_\kappa(\theta_1)+G_\kappa(\theta_2))\log\frac{L}{y}+O(1)\,.
\end{equation}
The $O(1)$ terms come from the crossover region. We have ignored them since we are interested in the two logarithmic terms.

\begin{figure}
    \centering
    \includegraphics[width=0.7\linewidth]{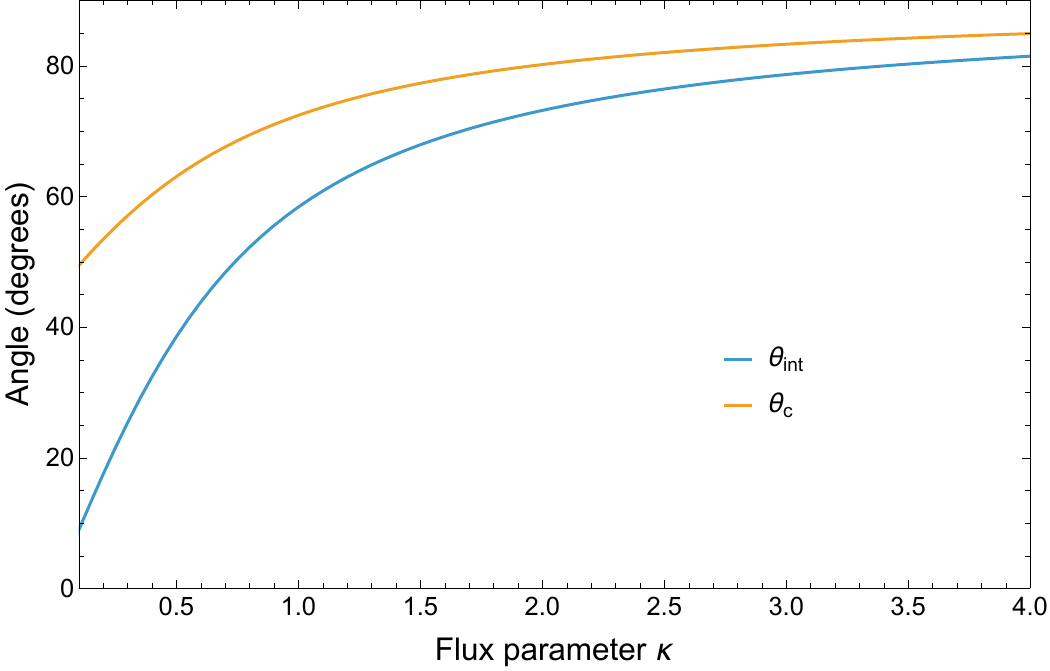}
    \caption{A plot of the threshold angle $\theta_{\rm int}$ below which the worldsheet in the vacuum saddle intersects the D5-brane, compared with the angle $\theta_c$ at which the vacuum saddle and the D5-ending saddle exchange dominance, for a range of the flux parameter $\kappa$. Importantly, the plot illustrates that $\theta_c>\theta_{\rm int}$ so both the saddles exist independently across the transition.}
    \label{fig:anglesvsflux}
\end{figure}

Note that (\ref{eq:D5freeenergy}) is not the final expression for the free energy. There is a competing saddle described in the bulk by a worldsheet that does not end on the D5-brane. It is described by a smooth worldsheet that originates on a defect ray, turns into the bulk and terminates on the other defect ray, and we refer to it as the vacuum phase since it is the only phase that contributes in the absence of the D5-brane. The competition between these two classical string saddles leads to a Gross–Ooguri-type transition \cite{GrossOoguri1998,Zarembo1999WilsonLoopCorrelator}. Closely related transitions involving fundamental strings ending on the probe D5-brane have previously been found for antiparallel and circular Wilson loops in the D3–D5 defect CFT \cite{PretiTrancanelliVescovi2017,BonanseaDavoliGriguoloSeminara2020}. The free energy of the vacuum phase is already computed in \cite{Drukker:1999zq} and we review it using our notation.
We use a scale invariant ansatz of the form (\ref{eq:embed}) but now there is a turning point in the bulk at the bisecting angle between the defect rays. For the vacuum saddle, it is convenient to shift the worldsheet angular coordinate $\phi$ so that the right defect-ray is at $\phi=0$ and left defect-ray is at $\phi=\chi$. The turning point is at $\phi=\frac{\chi}{2}$,
\begin{equation}
    h(\frac{\chi}{2})=h_m, \quad h'(\frac{\chi}{2})=0, \quad \chi\equiv \pi-\theta_1-\theta_2\,.
\end{equation}
$\chi$ is the opening angle at the cusp. The above condition just follows from symmetry since $h$ has to vanish along the two defect rays at the boundary.
The integration constant denoted $C_v$ (to distinguish from $C$ for the D5-ending case) is related to $h_m$ by
\begin{equation}
    C_v=\frac{\sqrt{1+h_m^2}}{h_m^2}\,.
\end{equation}
$h_m$ is determined in terms of the opening angle $\chi$ by\footnote{Depending on the cusp angles, the worldsheet could intersect the D5-brane before terminating. The condition for intersection is obtained by solving the D5 embedding condition using the solution derived above for the worldsheet,
\begin{equation}
    y+R\sin(\theta_2+\phi)=\kappa R h(\phi) \implies R_{\rm int}(\phi)=\frac{y}{\kappa h(\phi)-\sin(\theta_2+\phi)}\,.
\end{equation}
We are measuring the worldsheet angular coordinate $\phi$ counterclockwise from the right defect ray which is at an angle of $\theta_2$ with the x-axis.
The above intersection locus exists provided $R>0$ for some $0\leq\phi \leq \chi$, equivalently if
\begin{equation}
    \kappa > \text{min}\left(\frac{\sin(\theta_2+\phi)}{h(\phi)}\right)\,.
\end{equation}
For the symmetric cusp, when $\theta_1=\theta_2$, by symmetry, the minimum occurs along the angular bisector. In this case, the intersection occurs whenever the cusp angle is lower than $\theta_{\rm int}(\kappa)$ given by
\begin{equation} \label{eq:intangle}
    \kappa^{-1}=h(\frac{\pi}{2}-\theta_{\rm int})\,.
\end{equation}
See figure \ref{fig:anglesvsflux} for a plot of $\theta_{\rm int}$ against $\kappa$ by solving the above equation.}
\begin{equation}
    \frac{\chi}{2}=\int_0^{h_m}dh \frac{C_vh^2}{\sqrt{1+h^2}\sqrt{1+h^2-C_v^2 h^4}}\,.
\end{equation}
The renormalized action gives the vacuum cusp anomalous dimension as computed by \cite{Drukker:1999zq},
\begin{equation} \label{eq:vaccusp}
    \Gamma(\chi)=\frac{\sqrt{\lambda}}{2\pi}\left[-\frac{2}{h_m}+2\int_0^{h_m}dh \frac{C_v^2 h^2}{\sqrt{Q(h)}(\sqrt{1+h^2}+\sqrt{Q(h)})}\right], \qquad Q(h)\equiv 1+h^2-C_v^2 h^4.
\end{equation}
It can also be evaluated in closed form in terms of complete Elliptic integrals E and K matching with \cite{Drukker:2011za}. So, the free energy of this phase can be expressed entirely in terms of $\Gamma(\chi)$ as
\begin{equation}
    F_{\rm vac}=\Gamma(\pi-\theta_1-\theta_2)\log\frac{y}{a}+\Gamma(\pi-\theta_1-\theta_2)\log\frac{L}{y}+O(1)\,.
\end{equation}
Even though the two logarithms can be combined, we have written it in the above form to compare with the free energy of the D5-ending phase (\ref{eq:D5freeenergy}). Therefore, for given cusp angles $\theta_{1,2}$, the phase with lower free energy i.e., the one with smaller coefficient of the IR cusp logarithm will dominate,
\begin{equation}
    F=\text{min}\{F_{D5}, F_{\rm vac}\}\,.
\end{equation}
In the subsequent analysis, we will make the saddle comparison at the leading logarithmic order $\log\frac{L}{y} \gg 1$. Although the $O(1)$ terms could shift the transition locus, the shift is subleading at large $\frac{L}{y}$.
Comparing the logarithmic coefficients, the transition between the two phases occurs along the locus
\begin{equation}
    \Gamma(\pi-\theta_1-\theta_2)=\frac{\sqrt{\lambda}}{2\pi}(G_\kappa(\theta_1)+G_\kappa(\theta_2))\,.
\end{equation}
For the symmetric cusp, when $\theta_1=\theta_2$, the above transition occurs at the critical angle $\theta_c(\kappa)$,
\begin{equation}
    \Gamma(\pi -2\theta_c)=\frac{\sqrt{\lambda}}{\pi} G_{\kappa}(\theta_c)\,.
\end{equation}
For $\theta<\theta_c$, the D5-ending saddle dominates and for $\theta>\theta_c$, the vacuum saddle dominates. Intuitively, this is expected since for large angles, the defect rays are closer to each other, hence the interaction between them dominates their interaction with the interface. Also note that $\theta_c>\theta_{\rm int}$ as illustrated by the plot in figure \ref{fig:anglesvsflux}, so when we reduce the cusp angle from $\frac{\pi}{2}$, the transition occurs before the geometric intersection of the worldsheet with the D5-brane. This ensures that there is a genuine first-order phase transition that occurs at an angle at which the two phases exist independently.

\subsection{An aside: Relation to entanglement entropy across a wedge in AdS$_4$/BCFT$_3$}

Now, we comment on an interesting parallel between the calculation of the IR cusp logarithm in our setup with the calculation of entanglement entropy across a wedge in AdS$_4$/BCFT$_3$ \cite{Seminara:2017hhh}\footnote{For related work on corner contributions to entanglement entropy, including their holographic structure and universal relations to CFT data, see \cite{Bueno:2015rda, Bueno:2015xda}.}. They consider an entangling region on the $t=0$ spatial half-plane whose entangling curve is an oblique ray ending on the BCFT boundary at an angle $\gamma$. The corresponding Ryu-Takayanagi surface \cite{Ryu:2006bv} in the bulk is a two-dimensional surface ending orthogonally on the end-of-the-world brane sourced by the BCFT boundary condition \cite{Takayanagi:2011zk, Fujita:2011fp}. They parametrise the tension of the brane in terms of an angle $\alpha$. This is exactly analogous to our setup if we consider a single defect ray and take the $y\to 0$ limit so that the defect ray ends on the interface. The geometry of the resulting string worldsheet is identical to that of the RT surface with the following identifications between the parameters,
\begin{equation}
         \kappa_{\rm here} =- \cot \alpha_{\rm there}, \qquad
         \theta_{\rm here} = \gamma_{\rm there}, \qquad 
        \,.
\end{equation}
Since we restrict to $\kappa>0$, the correspondence covers the branch $\alpha \in [\frac{\pi}{2},\pi)$ in the AdS/BCFT parameter space.
The embedding functions for the worldsheet and the RT surface are related by
\begin{equation}
    h(\phi)_{\rm here}=\frac{1}{q(\phi)_{\rm there}}
\end{equation}
As a result, the renormalized area of the worldsheet which we expressed in terms of the function $G_\kappa(\theta)$ matches with the renormalized area of the RT surface which \cite{Seminara:2017hhh} parametrises in terms of $F_\alpha(\gamma)$. Precisely,
\begin{equation}
     G_\kappa(\theta)_{\rm here} = -F_\alpha(\gamma)_{\rm there}\,.
\end{equation}
The analogous observation for the AdS$_4$/CFT$_3$ case where the expression for the entanglement entropy across a wedge in the vacuum matches with (upto theory-dependent normalization factors) the negative of the holographic expression for the vacuum cusp anomalous dimension $\Gamma(\chi)$ in \cite{Drukker:1999zq}, was made in \cite{Hirata:2006jx}. \cite{Seminara:2017hhh} also provides a closed form expression for $F_\alpha(\gamma)$ in terms of incomplete Elliptic integrals, and in addition, observe that $F'_\alpha(\gamma)\leq 0$ and $F''_\alpha(\gamma)>0$. Our numerical observation (\ref{eq:completemonotonicity}) suggests complete monotonicity,
\begin{equation}
    (-1)^n F^{(n)}_\alpha(\gamma) \geq 0\,,
\end{equation}
of the universal boundary-corner function in the entanglement entropy. To our knowledge, this higher-derivative hierarchy has not been written down in the literature.

In addition, \cite{Seminara:2017hhh} also considers the case where the entangling surface is a wedge with the two rays of the wedge meeting on the BCFT boundary and observe a phase transition in the contribution to the holographic entanglement entropy between the RT surface connected to the end-of-the-world brane and the one which is disconnected from it. For the symmetric wedge with opening angle $\omega$, \cite{Seminara:2017hhh} writes the following expression for the corner function,
\begin{equation} \label{eq:enttransition}
    F_{\alpha, \rm wedge}=\text{max}\{F_e(\omega), 2F_\alpha(\gamma)\}, \qquad \omega=\pi-2\gamma.
\end{equation}
With the identification $G_\kappa=-F_\alpha$ and a similar identification between the vacuum corner function with the vacuum cusp anomalous dimension $\Gamma=-F_e$, the transition (\ref{eq:enttransition}) maps to the worldsheet transition in our setup.
In conclusion, after stripping off the overall theory-dependent normalizations, the free energy of the cusped Wilson line in the holographic D3-D5 setup matches with the logarithmic corner contribution to the entanglement entropy across a wedge in AdS$_4$/BCFT$_3$. We expect the parallel to also generalize to the finite-$y$ case with the logarithmic terms in the entanglement entropy taking a form similar to the free energy in (\ref{eq:D5freeenergy}).

\subsection{The generalized cusp with a tilt in the scalar polarization}

In the previous analysis, we set the scalar polarization of both the defect rays to be the same and to be along the $S^2$ wrapped by the D5-brane. We now allow the scalar polarization of each defect ray to tilt away from the $S^2$ wrapped by the D5-brane. For simplicity, we restrict the relevant $S^5$ motion to a common great-circle direction, so that the worldsheet is embedded in an AdS$_3\times S^1$ subspace of AdS$_5 \times S^5$. The metric on this subspace is
\begin{equation}
    ds^2=\frac{dz^2 + dx^2 + dx_\perp^2}{z^2}+d\zeta^2\,.
\end{equation}
$\zeta$ is a direction along the transverse $S^1$. The D5-brane is chosen to be at $\zeta=0$\footnote{The metric on the transverse $S^5$ can be written as
\begin{equation}
    ds^2_{S^5}=d\zeta^2 + \cos^2\zeta d\Omega_2^2+ \sin^2 \zeta d\Tilde{\Omega}_2^2\,.
\end{equation}
The D5-brane wraps the $S^2$ of unit radius at $\zeta=0$.
}. To compute the coefficient of the IR cusp logarithm using the D5-ending saddle, we again use a scale invariant ansatz for the worldsheet embedding,
\begin{equation}
    x=R\cos(\phi), \quad x_\perp=R\sin(\phi), \quad z=Rh(\phi), \quad \zeta=\zeta(\phi)\,.
\end{equation}
The boundary values of the two embedding functions at the AdS boundary and the D5-brane are, respectively,
\begin{equation}
    h(\theta)=0,\quad \zeta(\theta)=\alpha; \qquad h(\phi_0)=h_0, \quad \zeta(\phi_0)=0\,.
\end{equation}
subject to the boundary conditions for the worldsheet at the D5-brane,
\begin{equation}
    \sin(\phi_0)=\kappa h_0, \qquad h'(\phi_0)=-\frac{\kappa(1+h_0^2)}{\cos \phi_0}\,.
\end{equation}
Here, $\alpha \in [0,\frac{\pi}{2}]$ is the tilt in the scalar polarization relative to the D5-brane.
The Nambu-Goto action computed from the induced metric on the worldsheet takes the form,
\begin{equation}
    S=\frac{\sqrt{\lambda}}{2\pi}\int \frac{dR}{R}\int_\theta^{\phi_0} d\phi \mathcal{L}_{h,\zeta}, \qquad \mathcal{L}_{h,\zeta}\equiv \frac{\sqrt{1+h^2+h'^2+h^2(1+h^2)\zeta'^2}}{h^2}
\end{equation}
Observe that when $\zeta'=0$, we recover (\ref{eq:defectarmaction}) as expected. The above Lagrangian has two conserved quantities,
\begin{equation}
    \begin{split}
        & J\equiv \frac{\partial \mathcal{L}}{\partial \zeta'}=\frac{(1+h^2)\zeta'}{\sqrt{1+h^2+h'^2+h^2(1+h^2)\zeta'^2}},\\
        & C\equiv \mathcal{L}-\zeta'\frac{\partial \mathcal{L}}{\partial \zeta'}-h'\frac{\partial \mathcal{L}}{\partial h'}=\frac{1+h^2}{h^2\sqrt{1+h^2+h'^2+h^2(1+h^2)\zeta'^2}}\,,
    \end{split}
\end{equation}
arising respectively from the fact that the Lagrangian does not depend on $\zeta$ and has no explicit $\phi$ dependence. So, $J$ can be interpreted as the conserved momentum along the $S^5$ direction, while $C$ is the conserved Hamiltonian associated with translations in $\phi$. From the above equations, we see that it is convenient to define $q\equiv \frac{J}{C}$ to integrate the equations of motion from the D5-brane to the AdS boundary giving,
\begin{equation}
\begin{split}
    & \theta =\sin^{-1}(\kappa h_0)+\int_0^{h_0} dh \frac{C h^2}{\sqrt{1+h^2}\sqrt{Q(h)}},\\
    & \alpha = \int_0^{h_0} dh \frac{Cq}{\sqrt{1+h^2}\sqrt{Q(h)}}\,.
    \end{split}
\end{equation}
where $Q(h)\equiv 1+(1-C^2 q^2)h^2-C^2h^4$.
These equations determine the two unknowns $h_0$ and $q$. Note that $C$ is determined in terms of these using the D5-endpoint conditions,
\begin{equation}
    C=\frac{\cos \phi_0\sqrt{1+h_0^2}}{h_0^2\sqrt{1+\kappa^2+\frac{q^2 \cos^2\phi_0}{h_0^2}}}\,.
\end{equation}
When $q=0$, we see that these relations reduce to the previous case with D5-aligned polarization. The renormalized on-shell action gives the generalization of $G_\kappa(\theta)$ to 
\begin{equation}
    G_\kappa(\theta,\alpha)=-\frac{1}{h_0}+\int_0^{h_0}dh \frac{C^2(q^2+h^2)}{\sqrt{Q(h)}(\sqrt{1+h^2}+\sqrt{Q(h)})}\,.
\end{equation}
As a consistency check, we see that when $\alpha=0$, we recover $G_\kappa(\theta)$. Interestingly, numerical tests suggest that the hierarchy of alternating-sign of derivatives persists for non-zero $\alpha$,
\begin{equation}
    (-1)^{n+1}\frac{\partial^n G_\kappa(\theta,\alpha)}{\partial \theta^n}>0,\qquad n\geq 1\,,
\end{equation}
as illustrated for some chosen values of the parameters in (\ref{tab:holographic-complete-monotonicity}).
At large $\kappa$, the leading term takes the form,
\begin{equation}
    G_\kappa(\theta,\alpha)=-\frac{\kappa \cos\alpha}{\sin \theta}+O(\kappa^{-1})\,,
\end{equation}
which can again be reproduced using the D3-D5 fuzzy-funnel solution in the double-scaling limit of \cite{NagasakiTanidaYamaguchi2012} from the gauge theory at tree level for an oblique defect ray, 
\begin{equation} \label{eq:fuzzyfunnel}
    F_{\rm tree}(\theta,\alpha)=-\log \langle W \rangle_{\rm tree}=-\frac{k-1}{2}\frac{\cos\alpha}{\sin\theta}\log\frac{L}{y}\,.
\end{equation}
Interestingly, there is a locus in the $(\theta,\alpha)$ space where $G_\kappa(\theta, \alpha)$ takes a particularly simple form,
\begin{equation} \label{eq:BPSD5locus}
    G_\kappa(\theta, \frac{\pi}{2}-\theta)=-\kappa\,.
\end{equation}
This can be easily seen analytically by setting $\theta+\alpha=\frac{\pi}{2}$ in the above equations of motion for the worldsheet.

\begin{table}
\centering
\scriptsize
\setlength{\tabcolsep}{4pt}

\textbf{(a) vacuum saddle: derivatives in the opening angle $\chi$}

\vspace{2mm}

\resizebox{\textwidth}{!}{%
\begin{tabular}{c c c c c c c c c}
\hline
$(\theta,\psi)$ & $\chi$
& $C_1$ & $C_2$ & $C_3$ & $C_4$ & $C_5$ & $C_6$ & $C_7$ \\
\hline
$(0.20,\,0.50)$ & $0.490$
& $6.32$ & $24.6$ & $149$ & $1.22{\times}10^{3}$
& $1.25{\times}10^{4}$ & $1.53{\times}10^{5}$
& $2.18{\times}10^{6}$ \\

$(0.25,\,2.70)$ & $2.48$
& $0.224$ & $0.378$ & $0.198$ & $0.427$
& $0.681$ & $1.87$ & $4.86$ \\

$(0.60,\,1.10)$ & $0.892$
& $2.08$ & $4.21$ & $13.5$ & $60.8$
& $341$ & $2.30{\times}10^{3}$ & $1.80{\times}10^{4}$ \\

$(0.90,\,2.20)$ & $1.17$
& $1.27$ & $1.95$ & $4.51$ & $15.3$
& $65.4$ & $335$ & $2.00{\times}10^{3}$ \\

$(1.25,\,1.50)$ & $0.433$
& $7.99$ & $35.4$ & $244$ & $2.26{\times}10^{3}$
& $2.60{\times}10^{4}$ & $3.61{\times}10^{5}$
& $5.83{\times}10^{6}$ \\
\hline
\end{tabular}%
}

\vspace{4mm}

\textbf{(b) vacuum saddle: derivatives in the azimuthal angle $\psi$}

\vspace{2mm}

\resizebox{\textwidth}{!}{%
\begin{tabular}{c c c c c c c c}
\hline
$(\theta,\psi)$
& $C_1$ & $C_2$ & $C_3$ & $C_4$ & $C_5$ & $C_6$ & $C_7$ \\
\hline
$(0.20,\,0.50)$
& $6.18$ & $23.6$ & $140$ & $1.12{\times}10^{3}$
& $1.12{\times}10^{4}$ & $1.35{\times}10^{5}$
& $1.89{\times}10^{6}$ \\

$(0.25,\,2.70)$
& $0.146$ & $0.347$ & $0.120$ & $0.323$
& $0.371$ & $1.15$ & $2.33$ \\

$(0.60,\,1.10)$
& $1.62$ & $2.76$ & $7.08$ & $25.8$
& $118$ & $642$ & $4.09{\times}10^{3}$ \\

$(0.90,\,2.20)$
& $0.429$ & $0.596$ & $0.537$ & $1.11$
& $2.36$ & $6.71$ & $21.0$ \\

$(1.25,\,1.50)$
& $1.89$ & $2.81$ & $5.32$ & $14.4$
& $47.9$ & $192$ & $895$ \\
\hline
\end{tabular}%
}

\vspace{4mm}

\textbf{(c) D5-ending saddle: derivatives of
$G_\kappa(\theta,\alpha)$ at $\kappa=1$}

\vspace{2mm}

\resizebox{\textwidth}{!}{%
\begin{tabular}{c c c c c c c c}
\hline
$(\theta,\alpha)$
& $C_1$ & $C_2$ & $C_3$ & $C_4$ & $C_5$ & $C_6$ & $C_7$ \\
\hline
$(0.25,\,0)$
& $18.2$ & $146$ & $1.75{\times}10^{3}$
& $2.80{\times}10^{4}$ & $5.60{\times}10^{5}$
& $1.35{\times}10^{7}$ & $3.77{\times}10^{8}$ \\

$(1.10,\,0)$
& $0.712$ & $2.07$ & $4.23$ & $17.6$
& $75.2$ & $427$ & $2.66{\times}10^{3}$ \\

$(0.45,\,0.30)$
& $5.30$ & $24.0$ & $159$ & $1.41{\times}10^{3}$
& $1.57{\times}10^{4}$ & $2.09{\times}10^{5}$
& $3.26{\times}10^{6}$ \\

$(0.65,\,0.50)$
& $2.28$ & $7.43$ & $33.1$ & $206$
& $1.58{\times}10^{3}$ & $1.46{\times}10^{4}$
& $1.57{\times}10^{5}$  \\

$(1.05,\,0.70)$
& $0.646$ & $1.79$ & $3.93$ & $16.6$
& $75.7$ & $444$ & $2.93{\times}10^{3}$ \\
\hline
\end{tabular}%
}

\caption{
The table illustrates the alternating-sign derivative
hierarchy for the cusp logarithms in the holographic D3-D5 example.
For the vacuum cusp we remove the positive prefactor $\frac{\sqrt{\lambda}}{2\pi}$.
In (a),
$C_n\equiv(-1)^{n+1}\partial_\chi^n\Gamma(\chi)$,
where
$\cos\chi=\cos^2\theta\cos\psi+\sin^2\theta$.
In (b),
$C_n\equiv(-1)^{n+1}\partial_\psi^n
\Gamma(\chi(\theta,\psi))$
at fixed $\theta$.
In (c),
$C_n\equiv(-1)^{n+1}
\partial_\theta^n G_\kappa(\theta,\alpha)$
at fixed $\alpha$, for the choice $\kappa=1$.
All angles are given in radians.
Notice that all the displayed signed derivatives are positive.
}
\label{tab:holographic-complete-monotonicity}
\end{table}

\begin{figure}
    \centering
    \includegraphics[width=0.7\linewidth]{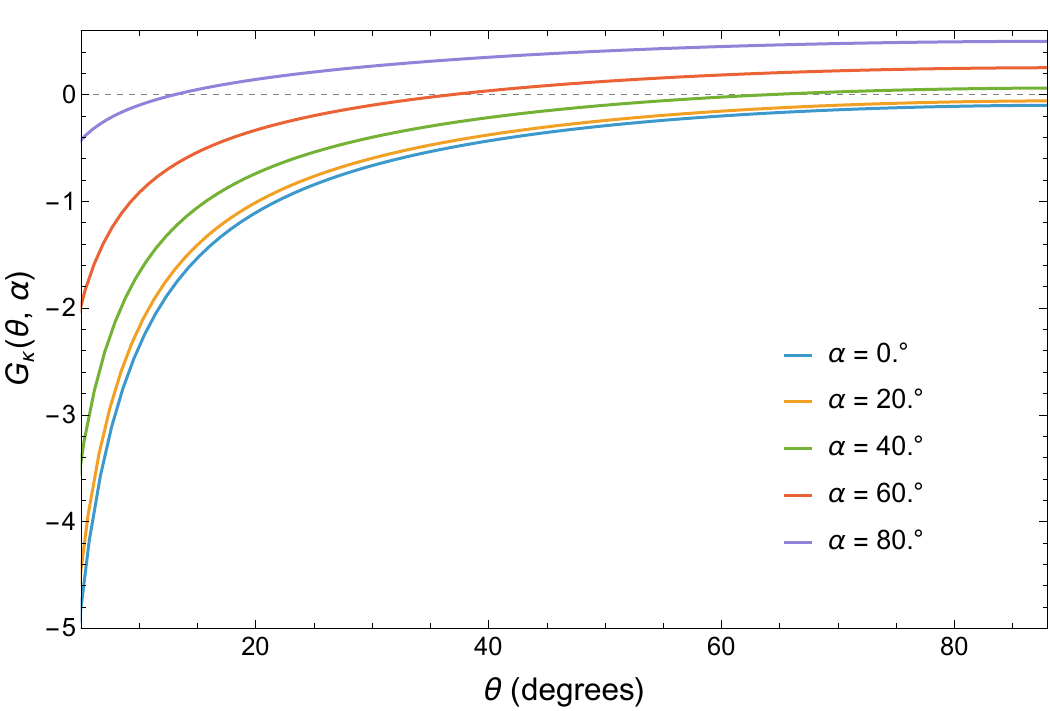}
    \caption{This is a graph of the function $G_\kappa(\theta,\alpha)$ governing the free energy of the generalized cusp in the D5-ending phase, against the geometric angle $\theta$ for various values of the scalar polarization angle $\alpha$. The flux parameter is chosen to be $\kappa=0.1$ for this plot. Observe that as we increase the value of $\alpha$, the function could be positive for large-enough geometric angles.}
    \label{fig:GeneralizedcuspD5}
\end{figure}

The calculation of the action for the vacuum phase also proceeds similarly so we skip the details. It reproduces the expression for the cusp anomalous dimension for the generalized cusp $\Gamma(\chi, \vartheta)$ computed in \cite{Drukker:2011za}. Here, $\chi$ is the geometric opening angle of the cusp and $\vartheta$ is the relative tilt in the scalar polarizations of the two defect rays. Summing up, the free energies of the D5-ending and vacuum phases takes the form,
\begin{equation} \label{eq:freeenergygeometricinternal}
    \begin{split}
        F_{D5} = & \Gamma(\chi, \vartheta) \log\frac{y}{a} + \frac{\sqrt{\lambda}}{2\pi}(G_\kappa(\theta_1,\alpha_1)+G_\kappa(\theta_2,\alpha_2))\log\frac{L}{y}+O(1),\\
        F_{\rm vac} = & \Gamma(\chi, \vartheta) \log\frac{y}{a} + \Gamma(\chi, \vartheta) \log\frac{L}{y}+O(1)\,.
    \end{split}
\end{equation}
Here $(\theta_1,\alpha_1)$ and $(\theta_2,\alpha_2)$ are the geometric and scalar
angles of the two defect rays, with $\chi=\pi-\theta_1-\theta_2,
    \quad
    \cos\vartheta=n_1^I n_2^I ,$
where $n_i^I$ are the unit scalar-polarization vectors. For the common
wrapped-$S^2$ orientation considered above, this reduces to
$\vartheta=|\alpha_1-\alpha_2|$.

Let us finally note that the special locus in (\ref{eq:BPSD5locus}) has a supersymmetric
interpretation. For the Maldacena-Wilson line (\ref{eq:MaldacenaWilson}), setting its supersymmetry
variation to zero gives the standard condition
\cite{Zarembo2002Supersymmetric}, which may be written as
\begin{equation}
    \Pi_i\epsilon=\epsilon,
    \qquad
    \Pi_i=\frac{1}{2}
    \left(1+i\,u_i^\mu\Gamma_\mu\,n_i^I\rho_I\right).
\end{equation}
Here $\epsilon$ is a constant Poincare supersymmetry spinor, $u_i^\mu$ is
the oriented unit tangent to the $i$th ray, and $\Gamma_\mu$ and $\rho_I$
denote gamma matrices along spacetime and the six scalar directions,
respectively. The D3-D5 defect imposes in addition the D5-brane
$\kappa$-symmetry projection \cite{Skenderis:2002vf}, which, translated to
$SO(4)\times SO(6)_R$ notation and choosing the D5 orientation used here,
can be written as
\begin{equation}
    \Pi_{\rm D5}\epsilon=\epsilon,
    \qquad
    \Pi_{\rm D5}
    =\frac{1}{2}
    \left(1+\Gamma_\perp\rho_1\rho_2\rho_3\right),
\end{equation}
where $\Gamma_\perp$ is associated with the direction normal to the
interface denoted $x_\perp$ in our notation and $\rho_{1,2,3}$ span the three internal directions whose unit
$S^2$ is wrapped by the D5-brane. For the planar cusp (4.19), the oriented
tangents obey
\begin{equation}
    u_1^\mu\Gamma_\mu
    =\cos\theta_1\,\Gamma_x-\sin\theta_1\,\Gamma_\perp,
    \qquad
    u_2^\mu\Gamma_\mu
    =\cos\theta_2\,\Gamma_x+\sin\theta_2\,\Gamma_\perp .
\end{equation}
A convenient supersymmetric choice of scalar orientations is
\begin{equation}
    n_1^I=-\cos\alpha_1\,e^I+\sin\alpha_1\,f^I,
    \qquad
    n_2^I=\cos\alpha_2\,e^I+\sin\alpha_2\,f^I ,
\end{equation}
where $e^I$ is a unit direction on the $S^2$ wrapped by the D5-brane and
$f^I$ is an orthogonal internal direction. Thus the components of the two
polarizations along the wrapped $S^2$ are antipodal. Compatibility of the three
projectors gives
\begin{equation}
    \alpha_i+\theta_i=\frac{\pi}{2},
    \qquad i=1,2,
\end{equation}
and leaves two of the sixteen Poincare supercharges, so the generic cusp on
this locus is $1/8$-BPS. Moreover,
$\vartheta=\pi-(\alpha_1+\alpha_2)=\theta_1+\theta_2$, and hence
$\chi+\vartheta=\pi$, as required by the usual BPS condition for the vacuum
generalized cusp along which the cusp anomalous dimension vanishes \cite{Drukker:2011za}. For the symmetric cusp, this reduces to
$\alpha=\frac{\pi}{2}-\theta$, precisely the locus appearing in (\ref{eq:BPSD5locus}). For the straight line $\theta=0$, it was already observed in \cite{NagasakiTanidaYamaguchi2012} that at $\alpha=\frac{\pi}{2}$, the line is `mutually supersymmetric to the interface'.

\subsection{Positivity constraints}

Before verifying the positivity constraints, we will first generalize the calculation of the free energy for the planar cusp described earlier to the general non-planar cusp. This is straightforward since we only have to replace the argument of the vacuum cusp anomalous dimension by the angle $\chi$ (\ref{eq:chiangle}) between the two defect rays. Since the interaction of the defect rays with the D5-brane factorizes, there is no effect of an out-of-plane deformation. Therefore, the free energy of the two phases is given by (\ref{eq:freeenergygeometricinternal}). Since the positivity constraints are derived on the reflection-symmetric slice, we need to set the scalar polarization angles to be equal, $\alpha_1=\alpha_2 \equiv \alpha$. This means the relative scalar polarization angle $\vartheta$ is zero.
We have already shown in Sec \ref{sec:Freeenergyscale} that $\Gamma(\chi)$ satisfies all the positivity constraints given that it is monotonic and concave in $\chi$ \cite{Cuomo:2024psk}. So, we now have to verify whether the coefficient of the IR cusp logarithm
\begin{equation}
    A_{\rm IR}=\text{min}\{\Gamma(\chi), \frac{\sqrt{\lambda}}{2\pi}(G_\kappa(\theta_1,\alpha)+G_\kappa(\theta_2,\alpha))\}\,,
\end{equation}
satisfies the positivity constraints (\ref{eq:IRpositivity}). Away from the phase transition locus, the vacuum phase satisfies all the positivity constraints since it is given by the vacuum cusp anomalous dimension $\Gamma(\chi)$, and the D5-ending phase trivially satisfies all the positivity constraints since it factorizes between the two defect rays. Consequently, its mixed angular Hessian vanishes identically, and its IR coefficient is independent of the azimuthal angle $\psi$. So, the only non-trivial check of the positivity constraints is across the phase transition locus. To this end, we define 
\begin{equation}
    \Delta \equiv \frac{\sqrt{\lambda}}{2\pi}(G_\kappa(\theta_1,\alpha)+G_\kappa(\theta_2,\alpha))-\Gamma(\chi)\,,
\end{equation}
in terms of which
\begin{equation}
    A_{\rm IR}=\Gamma(\chi)+\Delta \Theta(-\Delta)\,.
\end{equation}
Monotonicity in the azimuthal angle $\psi$ across the transition follows from
\begin{equation}
    \partial_\psi A_{\rm IR}=\Theta(\Delta)\partial_\psi \Gamma(\chi)\geq 0\,.
\end{equation}
To verify the constraints on the second derivatives, we define the Hessian
\begin{equation}
    H\equiv \frac{\partial^2 A_{\rm IR}}{\partial \lambda_L \partial \lambda_R}\,
\end{equation}
using the notation for the angle vectors $\lambda_{L,R}$ introduced in (\ref{eq:anglevectors}). We show that $H \preceq 0$ on the reflection-symmetric slice $\lambda_L=\lambda_R$. Across the transition, the above Hessian takes a distributional form,
\begin{equation}
    H=\Theta(-\Delta)H_{D5}+\Theta(\Delta) H_{\rm vac}-\delta(\Delta) (\partial_{\lambda_L}\Delta)(\partial_{\lambda_R}\Delta)^T\,.
\end{equation}
Since the three terms are negative semi-definite on the reflection-symmetric slice $\lambda_L=\lambda_R$, we have thereby verified that $A_{\rm IR}$ satisfies all the positivity constraints.

\section{Discussion}

In this work, we derived non-perturbative constraints on the free energy of a cusped line defect in the presence of a boundary or interface from reflection positivity. For planar cusps these imply mixed concavity, while for general non-planar cusps they imply negative semi-definiteness of the mixed angular Hessian and, in rotationally invariant setups, monotonicity in the azimuthal angle. In scale-invariant configurations, these constraints apply independently to the UV and IR logarithmic coefficients, and we verified them in free scalar and Maxwell theories, weakly coupled planar $\mathcal N=4$ SYM, and the holographic D3-D5 defect CFT, including across the transition between the competing worldsheet saddles. 

In all the examples studied, we furthermore find numerical evidence that the negatives of the logarithmic coefficients with nontrivial azimuthal angle dependence are completely monotonic functions of the azimuthal angle, a substantially stronger property than that implied by the two-point reflection-positivity constraints. A straightforward extension of reflection positivity to higher moments does not appear to explain this hierarchy. Positivity constrains moments rather than connected cumulants and therefore leads, when expressed in terms of free-energy derivatives, to nonlinear inequalities involving products of lower cumulants, for example the inequality involving the fourth cumulant and the square of the variance $\kappa_4+2\kappa_2^2\geq0$, rather than a sign constraint on $\kappa_4$ itself. Since the connected cumulants relevant to the universal logarithmic terms scale linearly with the large logarithms, whereas products of lower cumulants generate higher powers of these logarithms and mix the UV and IR contributions, such higher-moment inequalities do not yield independent sign constraints on the higher derivatives of either logarithmic coefficient.

An interesting future direction is to compute the IR logarithmic coefficient of the cusp in other examples of interacting CFTs. Natural examples are the pinning-field defects in the $O(N)$ Wilson-Fisher fixed points and the $3d$ Ising CFT
\cite{Cuomo:2021LocalizedMagnetic,ParisenToldin:2016Pinning,Allais:2014MagneticDefect}, for which the vacuum cusp anomalous dimension has already been studied \cite{Cuomo:2024psk}. The boundary critical behavior of these theories is also well developed, with conformal boundary conditions studied using the boundary bootstrap and, more recently, fuzzy-sphere and fuzzy-hemisphere constructions \cite{Gliozzi:2015BoundaryBootstrap,Metlitski:2022Boundary,Zhou:2025IsingSurface,Dedushenko:2024IsingBCFT,Hu:2026BoundaryBootstrap,Feng:2026ONSurface}. Combining these developments could provide perturbative and non-perturbative results for the IR cusp coefficient in these models and thereby provide new tests of the positivity constraints derived in this paper.

Now, we discuss another interesting future direction in relation to the holographic cusp calculation in the D3-D5 setup. The IR cusp logarithm that we computed does not directly determine the displacement two-point function of the straight Wilson line, since the small-angle limit is not uniform with the large-distance limit used to define the IR logarithmic coefficient. A complementary direction is therefore to study quadratic fluctuations of the string worldsheet describing a straight Wilson line near the D5 interface, which compute its displacement two-point function at strong coupling and probe how defect fluctuations interact with the interface. A closely related calculation was recently performed for a Wilson line in a finite-temperature state, where worldsheet fluctuations determine the thermal displacement correlator and exhibit bouncing singularities \cite{Giombi:2026Bouncing}. It would also be interesting to understand the relation to situations in which the defect is not treated as a probe: for example, in \cite{Chandra:2026ElasticStiffness} the displacement two-point function is obtained including the leading gravitational backreaction of a heavy line defect, leading to qualitatively different dynamical behavior.

\section*{Acknowledgments}

I would like to specially thank Zohar Komargodski and Yifan Wang for valuable discussions and for sharing useful references at various stages of the project, and for comments on the draft. I also thank Ahmed Abdalla, Tom Hartman and Luca Iliesiu for helpful discussions. This work was supported in part by the Leinweber Institute for Theoretical Physics at UC Berkeley; and by the Department of Energy, Office of Science, Office of High Energy Physics through award DE-SC0025293 and QuantISED award DE-SC0019380. 

\bibliographystyle{ourbst}
\bibliography{ref.bib}

\end{document}